\documentclass[showpacs,amsmath,amssymb,twocolumn]{revtex4}
\usepackage{amssymb}
\usepackage[dvips]{graphicx}

\begin{document}

\title{Microscopic theory of OMAR based on kinetic equations for quantum spin correlations.}
\author{A.V. Shumilin$^1$}

\affiliation{$^1$ Ioffe Institute, 194021 St.-Petersburg, Russia}

\begin{abstract}
The correlation kinetic equation approach is developed that allows describing spin correlations in a material with hopping transport. The quantum nature of spin is taken into account. The approach is applied to the problem of the bipolaron mechanism of organic magnetoresistance (OMAR) in the limit of large Hubbard energy and small applied electric field. The spin relaxation that is important to magnetoresistance is considered to be due to hyperfine interaction with atomic nuclei. It is shown that the lineshape of magnetoresistance depends on short-range transport properties. Different model systems with identical hyperfine interaction but different statistics of electron hops lead to different lineshapes of magnetoresistance including the two empirical laws $H^2/(H^2 + H_0^2)$ and $H^2/(|H| + H_0)^2$ that are commonly used to fit experimental results.
\end{abstract}

\maketitle

\section{Introduction}

Hopping conductivity is one of the fundamental types of electron transport in solid-state materials. It exists when the electron wavefunctions are localized. The conductivity is achieved due to the acts of hopping when the electron hops between localized functions (sites) with different energies due to the emission or absorption of phonons. The conventional theory of hopping conductivity is closely related to doped semiconductors with compensation. It is based on the mean-field approximation and Miller-Abrahams resistor network, which follows from this approximation in weak applied electric field \cite{MA, Efr-Sh}. The drawback of the mean-field approximation is that it neglects correlations between occupation numbers of different sites.

One type of materials with hopping transport that is actively developed right now is the organic semiconductors. They are already widely applied in OLEDs \cite{OLEDs} and have other possible applications, for ex. in organic solar cells \cite{SolarCells}. These materials very often display an intriguing property that is called ``organic magnetoresistance'' or ``OMAR'' \cite{OMAR0,OMAR,sheng, TDNguen, Shakya, Yan, OMAR-rev}. It is quite a strong magnetoresistance observed in magnetic fields $10-100gs$ both at low and room temperatures. Although the qualitative explanations \cite{Prigodin,Bobbert} and semi-qualitative theories \cite{HF1,HF2,HF3,Chi1,Gao,Lu2017,Larabi}  of this phenomenon started to appear ten years ago, the detailed microscopic theory of OMAR is not yet developed. One of the reasons for it is the close relation of OMAR to non-equilibrium spin correlations. The magnetoresistance is equal to zero in the mean-field approximation \cite{AVS-MF} but re-appears when the correlations are included in the theory even in an oversimplified model \cite{AVS-cor}. The physical reason for OMAR is the dependence of the relaxation of spin correlations on the applied magnetic field. This relaxation is often associated with hyperfine interaction with atomic nuclei. With some simplification, it can be described as spin rotation around the so-called ``hyperfine fields''. These fields are different at different sites therefore random hopping of electron with rotation around these fields leads to spin relaxation. When the external magnetic field is large compared to hyperfine fields, the spin rotates around approximately the same direction on all the sites and its relaxation is suppressed.

The microscopic theory of OMAR requires a theoretical approach that takes into account the non-equilibrium correlations including the spin correlations. Up to very recent times, practically the only theoretical tool to do this was the Monte-Carlo numerical simulation. It was used in one of the pioneering studies of OMAR to show the possibility of its bipolaron mechanism (the mechanism related to double-occupation of a single site with two electrons in the spin-singlet state) \cite{Bobbert}.  However, this method has its drawbacks. It is a numerical method not suited for analytical theory. Also, it is based on the semi-classical nature of hopping transport were all the quantum mechanics is included in the electron hopping rates. It has some problems with spin correlations that actually have quantum nature. In \cite{Bobbert} the spin was described semi-classically as the two possibilities for an electron: to have spin up or spin down. This approximation can readily be used in Monte-Carlo simulation, however, it cannot describe actual spin rotation around hyperfine fields. To describe it the spin should be allowed to be directed along any axis, not only ``up'' and ``down''. It will be shown that it requires a more rigorous description of spin correlations based on quantum mechanics.

There is an approach to consider pair correlations in close pairs of sites as a modification of Miller-Abrahams resistors \cite{AVS-cor,Aleiner1,Aleiner2}. However, up to now, electron spin was considered with this approach only in the semi-classical ``up'' and ``down'' model.

Very recently the approach that allows to include correlations of arbitrary order into the analytical theory of hopping transport was developed \cite{CKE0,CKEres}. The approach operates with correlation kinetic equations (CKE) that relate occupation numbers and their correlations. It is based on Bogolubov chain of equations \cite{Bogolubov,Landau10}. In \cite{CKE0,CKEres} this approach is developed only for charge correlations, i.e. it does not consider electron spin in any model.

The goal of this study is to develop CKE theory that includes spin correlations and explicitly takes into account their quantum nature and to apply this theory to the problem of organic magnetoresistance. The study is restricted to the bipolaron mechanism of OMAR in low electric fields and large Hubbard energy. The spin relaxation is considered to be provided by hyperfine interaction with atomic nuclei.

Although the qualitative theories can give a general understanding of OMAR it is desirable to have an approach that can be used to calculate OMAR explicitly. The progress in general understanding of organic semiconductors and in simulation technics leads to the possibility to calculate the microscopic properties of organic materials: the energy of molecular orbitals and their overlap integrals \cite{Masse}. Some additional study of electron-phonon interaction in organic materials may lead to the possibility of direct calculation of hopping rates. If all these properties will be known the CKE approach can be used to quantitatively calculate the magnetoresistance. In the present paper, the magnetoresistance is calculated in model systems. It is shown that the so-called lineshape of magnetoresistance (the shape of the dependence of resistivity on the applied magnetic field) depends on the properties of short-range electron transport. Different model systems with identical hyperfine interaction but different statistics of hopping rates show different lineshapes of organic magnetoresistance. The obtained lineshapes include (but are not limited to) the two empirical laws that are most often used to fit experimental data \cite{OMAR,sheng}: $H^2/(H^2 + H_0^2)$ and $H^2/(|H| + H_0 )^2$. Here $H$ is the applied magnetic field and $H_0$ is a fitting parameter. It gives hope that the calculations of OMAR can be used to relate microscopic models of hopping transport in organics with experimental results.

The paper is organized as follows. In Sec.~\ref{s:mod} I discuss the model that is used to describe organic semiconductors. In Sec.~\ref{s:cor-def} the general mathematical definitions of quantum spin and charge correlations are introduced. In Sec.~\ref{s:CKE} the kinetic equations that relate these correlations to currents are derived. In Sec.~\ref{s:MR} the obtained system of CKE is used to describe the possible lineshapes of the bipolaron mechanism of OMAR. In Sec.~\ref{s:MR-an} it is done analytically in the model of modified Miller-Abrahams resistors. In Sec.~\ref{s:MR-num} it is done in the more general case with the numerical solution of CKE. In Sec.~\ref{s:dis} the general discussion of the obtained result is provided. In Sec.~\ref{s:conc} the conclusion is given. Some part of quantum mechanical calculations that are made to derive spin CKE is discussed in the appendix \ref{Ap1}.

\section{model}
\label{s:mod}

The following model is considered in the present work. The material contains a number of hopping sites were electrons (or polarons) are localized. In principle a hopping site can contain two polarons, however, the energy of its double occupation is larger than the energy of its single occupation. If the energy of the single occupation of site $i$ is $\varepsilon_i$, the energy of its double occupation is $\varepsilon_i + U_h$, were $U_h$ is the Hubbard energy. The spins of electrons on a double-occupied site should form spin-singlet. The possibility of double occupation with electrons in a triplet state is neglected. It is considered that the system has some concentration of electrons and the Fermi level $\mu$. The current in small applied electric fields in the linear response regime is discussed.

I consider the Hubbard energy to be much larger than temperature and energy differences in the hopping process. In this situation all the sites participating in hopping transport can be divided into the two groups. $A$-type sites have the energy of single-occupation near Fermi energy $\varepsilon_i \sim \mu$. They have 0 or 1 electrons but are never double-occupied because $\varepsilon_i + U_h$ is too large for an $A$-type site $i$. $B$-type sites have the energy of double occupation near chemical potential $\varepsilon_j + U_h \sim \mu$. They always have at least one electron because $\varepsilon_j \ll \mu$ for a $B$-type site $j$. Therefore ``unoccupied'' $B$-type site is a $B$-type site with one electron and has the spin degree of freedom. An occupied $B$-type site has two electrons in the spin-singlet state. This model is most convenient for the description of a situation when the distribution of site energies $\varepsilon_i$ is broad not only compared to temperature but also to the Hubbard energy $U_h$ (Fig. \ref{fig:AB} (A)). In this case, only a small part of hopping sites effectively participate in transport. The consideration of sites without an energy level near $\mu$ significantly complicates the numeric simulation and has little impact on the result. I adopt the simplified model where there are two independent densities of states for $A$-type sites and for $B$-type sites (Fig. \ref{fig:AB} (B)).

\begin{figure}[htbp]
    \centering
        \includegraphics[width=0.49\textwidth]{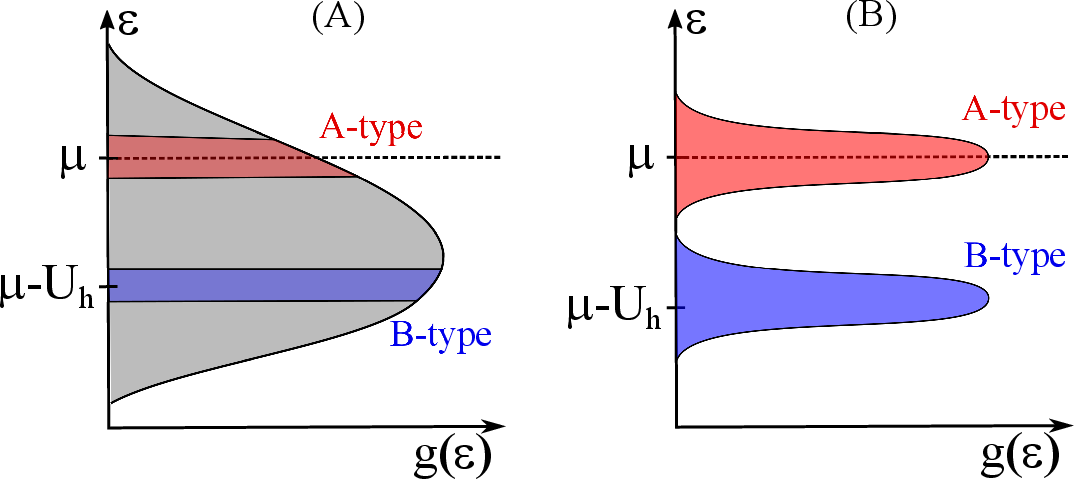}
        \caption{The density of states in the system with a broad distribution of energies (A) and in the adopted simplified model (B).
        The $A$-type and $B$-type sites marked with red and blue color correspondingly.  }
    \label{fig:AB}
\end{figure}

In the following part of the text, the energy of double occupation of $B$-type site $i$ is denoted as $\varepsilon_i$ because the energy of its single occupation does not appear in the theory.

The hopping between sites is controlled by hopping rates $W_{ij}$. It is assumed that the electron spin is always conserved in the hopping process. When sites $i$ and $j$ have the same type, the site $j$ is occupied and site $i$ is not, the electron hops from $j$ to $i$ with rate $W_{ij}$.
\begin{equation}\label{Wij}
W_{ij} = W_0 |t_{ij}|^2 \exp\left( - \frac{\max[(\varepsilon_i - \varepsilon_j), 0]}{T} \right).
\end{equation}
Here $t_{ij}$ is the overlap integral between localized states on sites $i$ and $j$. $W_0$ describes the strength of electron-phonon interaction. In principle, it can have a power-law dependence of site energies $\varepsilon_i$, $\varepsilon_j$. However, this dependence is not universal (is material-depend) and is not considered in the present study. Therefore $W_0$ is treated as a constant. When site $j$ is of type $B$ and $i$ is $A$-type site ($B\rightarrow A$ hop) the hopping occurs with the rate $2W_{ij}$ because both of the two electrons on site $j$ can hop to $i$. After the hop, the spins of electrons are in the singlet state. In the situation of $A \rightarrow B$ hop, when spins are in thermal equilibrium, the hopping occurs with the rate $W_{ij}/2$. However, this rate increases to $2W_{ij}$ when the spins are in singlet state.
It ensures that the detailed balance holds in the thermal equilibrium in a pair of sites of different kinds.
The $A \rightarrow B$ hop is impossible when the spins are in a triplet state because the electron spin is conserved during the hop.

The equilibrium occupation number of an $A$-type site $i$ is equal to $n_i^{(0)} = 1/(1+e^{(\varepsilon_i-\mu)/T}/2)$. The factor $2$ in this expression follows from the degeneracy of the occupied state of an $A$-type site \cite{Efr-Sh}. The $A$-type site $i$ has two occupied states with different spin. The joint probability of the occupied states is $2 e^{-(\varepsilon_i-\mu)/T}/Z_i$ where $Z_i$ is the statistical sum of site $i$:  $Z_i = 1 + 2e^{-(\varepsilon_i-\mu)/T}$. It leads to the mentioned expression for the occupation probability $n_i^{(0)}$.  For the $B$-type site $j$ the free state is degenerate. With similar arguments it leads to slightly different expression for the equilibrium  occupation number of site $j$: $n_j^{(0)} = 1/(1+ 2e^{(\varepsilon_j-\mu)/T})$. 

In any pair of sites $i-j$, the equality holds without respect for site types
\begin{multline}\label{Gamma}
\Gamma_{ij} = W_{ij}p_{sp}^{(ij)}(1-n_i^{(0)}) n_j^{(0)} =  \\
W_{ji} p_{sp}^{(ji)} (1-n_j^{(0)}) n_i^{(0)} = \Gamma_{ji}.
\end{multline}
$\Gamma_{ij}$ is the number of electrons that hops from site $j$ to site $i$ in unit of time in thermal equilibrium. $p_{sp}^{(ij)}$
is the spin term in the hopping probability. It is equal to $1/2$ when site $j$ has type $A$ and site $i$ has type $B$. When site $j$ is $B$-type site and $i$ is $A$-type site, $p_{sp}^{(ij)} = 2$. When the types of sites $i$ and $j$ are the same, $p_{sp}^{(ij)} = 1$.
 The equation (\ref{Gamma}) shows that detailed balance holds in thermal equilibrium.

\begin{figure}[htbp]
    \centering
        \includegraphics[width=0.35\textwidth]{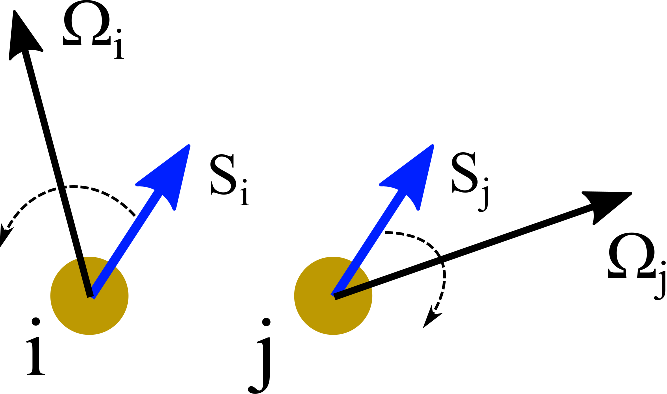}
        \caption{Rotation of spins on sites $i$ and $j$ with Larmor frequencies $\mathbf{\Omega}_i$ and $\mathbf{\Omega}_j$.  }
    \label{fig:AB}
\end{figure}

The organic magnetoresistance is closely related to the dynamics of spin correlations. The mathematical description of these correlations is introduced in Sec.~\ref{s:cor-def}. Here I describe the physics that is considered. The main reason for spin dynamics is the hyperfine interaction with atomic nuclei. It is described as effective on-site magnetic fields ${\bf H}_{\rm hf}^{(i)}$ that are different on different sites. These fields should be added to the external magnetic field ${\bf H}$. The electron spin rotates around the total field with Larmor frequency $\mathbf{\Omega}_i = \mu_b g ({\bf H} + {\bf H}_{\rm hf}^{(i)})$. This description is valid when the slow dynamics of nuclear spins can be neglected.

 The rotation in hyperfine fields can modify spin correlations. Consider that at some point of time the spins of electrons on sites $i$ and $j$ were parallel. After some time due to rotation with different vector frequencies $\mathbf{\Omega}_i$ and $\mathbf{\Omega}_j$ there will be an angle between spin directions. In combination with electron
hops, this rotation leads to spin relaxation \cite{Yu-HF,Harmon-Relax,HFine2}.

I also introduce a phenomenological time of spin relaxation $\tau_s$. It can describe the time of on-site spin relaxation related to (for example) the spin-phonon interaction. However, the main reason to introduce $\tau_s$ is the possibility to compare the results with previous theories where spin relaxation was treated in this way \cite{Bobbert,AVS-cor}.

\section{Spin correlations}
\label{s:cor-def}

The site occupation numbers are insufficient for the description of OMAR that is controlled by spin degrees of freedom. OMAR appears due to spin correlations. Starting from the first qualitative studies of OMAR \cite{Bobbert,Prigodin} the  magnetoresistance is attributed to different probabilities of parallel and anti-parallel configurations of two spins. In terms of statistics the description of these probabilities is possible only beyond the mean-field approximation. By definition the mean-field approximation deals with occupation numbers on a single site. It can describe only the situation when one spin direction on some site is more probable than another without respect to other sites. It corresponds to the appearance of averaged on-site spin polarization that is impossible at room temperatures and magnetic fields $\sim 100 gs$. OMAR is controlled by the probabilities of the relative directions of two spins. For example let us consider the situation when parallel configuration is more probable than anti-parallel. In this case the probability for both spins to have up direction $\uparrow\uparrow$ is equal to the probability of $\downarrow\downarrow$ direction but is larger than the probability of $\uparrow\downarrow$ and $\downarrow\uparrow$ directions.
In terms of statistics this situation is described with correlations. When the correlations are neglected in the mean-field approximation, it is impossible to distinguish between  parallel configuration of two spins (without averaged polarization) and the equilibrium statistics. Therefore a microscopic theory of OMAR should be developed beyond the mean-field approximation and take at least some correlations into account. The approach adopted in the present paper is to write the equations for arbitrary correlations and neglect the insignificant ones at the last step of the theory when the equations are solved numerically.

In this section, the general notations for spin correlations between electrons on different sites are introduced. The special attention is paid to the quantum nature of spin. It allows to describe simultaneously the hopping transport and the spin rotation around local hyperfine fields. These fields are responsible for spin relaxation in different materials \cite{Yu-HF,Harmon-Relax,HFine2} including many organic semiconductors.

 At first, I describe the electron state on a single A-type site $i$. The electron has the two quantum-mechanical states: with spin up $|\uparrow \rangle$ or down $|\downarrow \rangle$. The general description of its state can be given by $2 \times 2$ density matrix $\widehat{\rho}_i$. Its diagonal terms are real. Their sum is unity because the site $i$ is considered to be occupied. The non-diagonal terms are complex and are conjugate. It leads to three independent parameters describing the density matrix. These parameters can be selected to have clear physical meaning \cite{Bryksin-spin}: the averaged spin polarizations of site in the directions $x$, $y$ and $z$. The density matrix $\widehat{\rho}_i$ can be re-constructed with this averaged polarizations. $\widehat{\rho}_i = (1/2)\left( \widehat{1} + \overline{s}_{i}^x \widehat{\sigma}_x  + \overline{s}_{i}^y \widehat{\sigma}_y + \overline{s}_{i}^z \widehat{\sigma}_z\right)$.
Here $\overline{s}_{i}^{x,y,z}$ are the averaged values of operators $\hat{s}_{i}^{x,y,z}$ of spin polarization along the axes $x$, $y$ and $z$. $\widehat{\sigma}_{x,y,z}$ are the Pauli matrices. $\widehat{1}$ is the unit $2\times 2$ matrix. In this work it also will be denoted as $\widehat{\sigma}_0$

Now consider an $A$-type site $i$ with finite occupation probability $\overline{n}_i \equiv \overline{s}_i^0$. When the conductivity is due to electron hops the terms of density matrix with uncertain occupation numbers can be neglected between hops (although they should be treated with perturbation theory when the hopping rates are calculated). The density matrix $\widehat{\rho}_i$ can be given in block-diagonal form. The free site is described by the block $\widehat{\rho}_i^{(0)}$ that actually is a single number equal to $1-\overline{n}_i$.  The single-occupied site is described by the block $\widehat{\rho}_i^{(1)}$ that can be expressed as follows
\begin{multline}\label{rho1}
\widehat{\rho}_i^{(1)} = \frac{\sum_p \overline{s_i^p} \sigma_p}{2} = \\
 \frac{1}{2}\left(  \overline{s}_{i}^0 \widehat{\sigma}_0 + \overline{s}_{i}^x \widehat{\sigma}_x  + \overline{s}_{i}^y \widehat{\sigma}_y + \overline{s}_{i}^z \widehat{\sigma}_z\right).
\end{multline}
Here index $p$ can have one of the four values $(0,x,y,z)$. The averaged quantities $\overline{s}_{i}^0$, $\overline{s}_{i}^x$, $\overline{s}_{i}^y$ and $\overline{s}_{i}^z$ can describe any state of the site $i$.

It is possible to describe spin correlations in a similar manner.
The density matrix of two sites $i$ and $j$ can be expressed as the four blocks with well-defined occupation numbers. Let $i$ be an $A$-type site and $j$ be a $B$-type site. In this case the site $i$ can have $0$ or $1$ electrons and $j$ can have $1$ or $2$ ones. The joint density matrix of sites $i$ and $j$ can be expressed in terms of four blocks
\begin{equation}
\widehat{\rho}_{ij} = \left(
\begin{array}{cccc}
\widehat{\rho}_{ij}^{(12)} & 0 & 0 & 0 \\
0& \widehat{\rho}_{ij}^{(11)} &0 & 0 \\
0 & 0 & \widehat{\rho}_{ij}^{(02)} & 0 \\
0 & 0 & 0 & \widehat{\rho}_{ij}^{(01)}
\end{array}
\right).
\end{equation}
Here the upper indexes stand for the occupation numbers of the sites. For example, $\widehat{\rho}_{ij}^{(12)}$ describes the part of density matrix related to single-occupied site $i$ and double-occupied site $j$.
$\widehat{\rho}_{ij}^{(11)}$ is $4 \times 4$ matrix that describes the single-occupied state of both sites and contain all the spin correlations
\begin{equation}\label{rho2}
\widehat{\rho}_{ij}^{(11)} = \frac{1}{4}\sum_{pq} \overline{s^{p}_{i} s^{q}_{j}} \left(\widehat{\sigma}_{p}^{(i)} \otimes \widehat{\sigma}_{q}^{(j)}\right)
\end{equation}
where $\overline{s^{p}_{i} s^{q}_{j}}$ is the quantum mechanical average of operator $\hat{s}^{p}_{i} \hat{s}^{q}_{j}$.
Here $\hat{s}^{x,y,z}_{i}$ and $\hat{s}^{x,y,z}_{j}$ are the operators of spin polarizations of sites $i$ and $j$ correspondingly. $s^{0}_i$ and $s^{0}_j$ are the operators of single occupations of sites $i$ and $j$. They always have well-defined values between hops. Note that the relation between occupation number and $\overline{s}^{0}$ is different for different types of sites. For A-type site $i$, $\overline{s}^0_{i} = \overline{n}_i$. For a B-type site $j$, $\overline{s}^0_{j} = 1- \overline{n}_j$ because $B$-type sites have one electron in unoccupied state. I keep these double notations because the notation $n_i$ is useful to track the charge conservation law while the notation $\hat{s}^0$ allows to give the expression for density matrix in terms of correlations in unified form for both A-type and B-type sites.

The sign $\otimes$ in Eq.~(\ref{rho2}) denotes the Cartesian product of matrices. The upper indexes $(i)$ and $(j)$ in $\widehat{\sigma}_{p}^{(i)}$ and $\widehat{\sigma}_{q}^{(j)}$ have no mathematical meaning but help to track what Pauli matrix is related to what site.

The block $\widehat{\rho}_{ij}^{(02)}$ is a single number equal to $\overline{(1-n_i)n_j}$
The blocks $\widehat{\rho}_{ij}^{(12)}$ and $\widehat{\rho}_{ij}^{(01)}$  are $2\times 2$ matrices
\begin{multline}\label{rho2-1}
\widehat{\rho}_{ij}^{(12)} = \sum_p \overline{ (1 - s^0_{j}) s^{p}_{i}} \widehat{\sigma}_p^{(i)}, \\
\widehat{\rho}_{ij}^{(01)} =  \sum_p \overline{(1 - s^0_{i})s^{p}_{j}} \widehat{\sigma}_p^{(j)}.
\end{multline}

The whole matrix $\rho_{ij}$ can be parameterized with 24 averaged values: $\overline{s^{p}_{i}}$, $\overline{s^{p}_{j}}$ and
$\overline{s^{p}_{i}s^{q}_{j}}$. These values have clear physical meaning: they describe the occupation probabilities, mean spin polarizations and their correlations. The kinetic equations for these averaged values would allow to describe all the dynamics of the density matrix for a system with hopping transport. This result can be generalized for arbitrary number of sites. Let $I$ be some set of sites. The density matrix $\widehat{\rho}_I$ of this set can be described by averaged products of $\hat{s}_i^{p}$ in all the subsets $I_n$  of the set $I$.
\begin{equation}\label{sI}
\overline{s_{I_n}^{P}} = \overline{\prod_{i \in I_n} \hat{s}_{i}^{p_i}  }, \quad P = \{p_1,... p_{n(I_n)} \}, \quad I_n \subset I.
\end{equation}
Here $P$ is the set of upper indexes $p_i$ equal to $0$, $x$, $y$ or $z$ related to the sites $i$ in the set $I_n$.

All the kinetics of a hopping system can be described with $\overline{s_I^{P}}$. However, these values cannot be considered as correlations. When two A-type sites $i$ and $j$ are not correlated $\overline{s_i^{0} s_j^{0}} = \overline{n}_i \overline{n}_j \neq 0$ in thermal equilibrium. Therefore $\overline{s_I^{P}}$ cannot be neglected even when correlations inside the set $I$ are not important.

The idea of correlation kinetic approach \cite{CKE0} is to write equations for correlations themselves and neglect the correlations at large distance and of high order. To follow this idea the correlations $\overline{c_{I}^{P}}$ are introduced
\begin{equation}\label{cI}
\overline{c_{I}^{P}} = \overline{\prod_{i \in I} c_{i}^{p_i}  }, \quad
c_{i}^{0} = n_i - n_i^{(0)}, \quad c_i^{\alpha} = \hat{s}_i^{\alpha}.
\end{equation}
Here the Greek letter $\alpha$ stands for the spin projection on $x$, $y$ or $z$. $n_i^{(0)}$ is the equilibrium filling number. In this notations $\overline{c}_{i}^{\alpha} = 0$ when we neglect averaged spin polarization and $\overline{c}_{i}^{0} = \overline{n}_i - n_i^{(0)}$ is the perturbation of the occupation number due to applied electric field. When the hopping system is close to equilibrium, eq.~(\ref{cI}) describes the correlations that can be neglected if occupation numbers and spins in the set $I$ are not considered to be correlated.

\section{Kinetic equations for spin and charge correlations}
\label{s:CKE}

In this section the kinetic equations are derived for the correlations defined in Sec.~\ref{s:cor-def}. Consider the correlation $\overline{c}_I^P$ in some set $I$ of sites. It can be changed due to one of the following processes: the hopping of electrons between sites of the set and outer sites, the hopping of electrons inside set $I$ and due to the internal spin dynamics.
\begin{multline}\label{kin-gen-1}
\frac{d}{dt} \overline{c}_I^P = \sum_{i\in I, k \notin I} \left(\frac{d}{dt}\right)_{ik} \overline{c}_I^P  +  \\
 \sum_{i,j \in I} \left(\frac{d}{dt}\right)_{ij} \overline{c}_I^P
 + \sum_{i\in I} \left(\frac{d}{dt}\right)_{i} \overline{c}_I^P
\end{multline}
Here in r.h.s of Eq.~(\ref{kin-gen-1}) the symbolical expressions for different terms are used. $(d/dt)_{ij} \overline{c}_I^P$ means the changing rate of $\overline{c}_I^P$ due to the hops between sites $i$ and $j$ of the set. $(d/dt)_{ik} \overline{c}_I^P$ stands for the transitions between site $i$ from the set and outer site $k$. $(d/dt)_{i} \overline{c}_I^P$ describes the changing rate of $\overline{c}_I^P$ due to the internal spin dynamics (spin rotation and phenomenological spin relaxation) related to site $i$.

The term $(d/dt)_{ik}$ describes the transition of correlations between sets of sites. It can be expressed as follows.
\begin{equation}
\left( \frac{d}{dt} \right)_{ik} \overline{c_i^p c_{I'}^{P'}} = T_{ik}^{(p)} \overline{c_k^p c_{I'}^{P'}} - T_{ki}^{(p)}\overline{c_i^p c_{I'}^{P'}}.
\end{equation}
Here $T_{ik}^{(p)}$ is the rate of charge or spin transition from site $k$ to site $i$. $c_{I'}^{P'}$ describes the part of correlation that is not related to site $i$ and is conserved during $i \leftrightarrow k$ hops. $I' = I \backslash \{i\}$ where the notation $\backslash$ stands for the set difference.  $P'$ is the set of indexes from $P$ other than the index $p$ related to the site $i$.
\begin{equation} \label{c-tr}
T_{ik}^{(0)} = W_{ik}p_{sp}^{(ik)}\left(1-n_i^{(0)}\right) + W_{ki} p_{sp}^{(ki)}n_i^{(0)}
\end{equation}
$T_{ik}^{(0)}$ describes the rate of transition of small perturbation of charge density from site $k$ to site $i$ \cite{CKE0}.

\begin{figure}[htbp]
    \centering
        \includegraphics[width=0.35\textwidth]{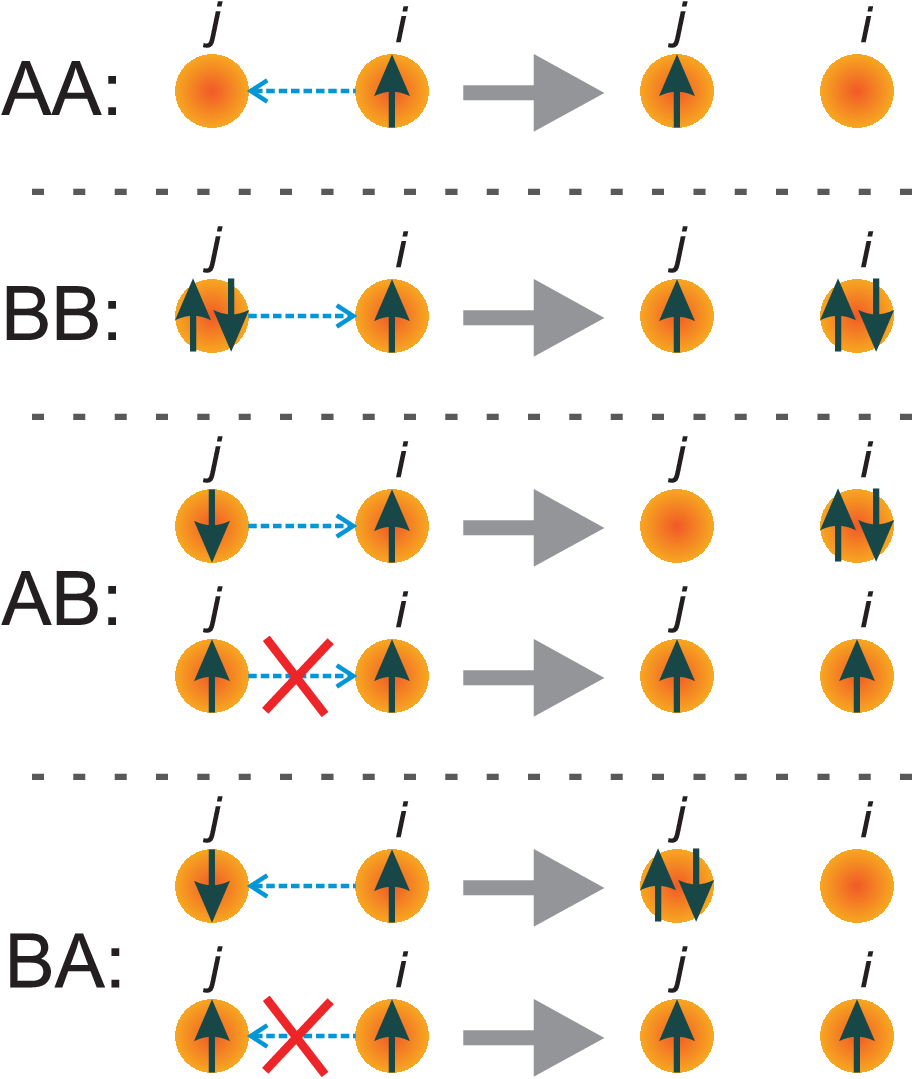}
        \caption{The spin transfer from site $i$ to site $j$ in different pairs of sites. On the left hand side of the figure the site $i$ is spin-polarized and site $j$ is not. On the right-hand side of the figure (after the hop) the site $j$ acquires spin polarization in the direction of the initial polarization of site $i$. }
    \label{fig:str}
\end{figure}

$T_{ik}^{(\alpha)}$ describes the process of spin polarization transfer between sites. This process is different for different types of sites, as shown in Fig. \ref{fig:str}. In a pair of A-type sites the spin transfer is achieved due to hops of spin-polarized electrons. In a pair of B-type sites the hops of spin-polarized holes are responsible for the spin transfer. In a mixed $AB$ pair the spin transfer occurs because electron with one spin projection can hop from the $A$-type site to the $B$-type site while the electron with other spin projection cannot. It leads to the following expressions for spin transfer rates in different pairs of sites (see \cite{AVS-MF} for details).
\begin{equation}
T_{ik}^{(\alpha)} = \left\{
\begin{array}{ll}
W_{ik}(1-n_i^{(0)}), & AA \\
W_{ki} n_i^{(0)}, &BB \\
W_{ki} n_i^{(0)}/2, &AB \\
W_{ik} (1-n_i^{(0)})/2, &BA
\end{array}
\right.
\end{equation}

The term $(d/dt)_{ij}$ for charge correlations is derived in \cite{CKE0} using the fact that joint occupation of sites $i$ and $j$ cannot be changed due to $i \leftrightarrow j$ hops, $(d/dt)_{ij} \overline{n_in_j} = 0$. This argument should be generalized to include spin and different types of sites. When both the sites $i$ and $j$ have the same type the hops between them are impossible when both of them are single occupied. It leads to the expression $(d/dt)_{ij} \overline{s_i^p s_j^q s_{I'}^{P'}} = 0$ for $AA$ and $BB$ pairs of sites $i$, $j$. When the sites $i$ and $j$  have different types the hop between them is possible when both are single-occupied. Actually it is the process that relates spin correlations and charge transport in bipolaron mechanism of OMAR
. Rate equations for this process when $i$ is $A$-type site and $j$ is $B$-type site are derived in appendix \ref{Ap1} with quantum mechanic approach:
\begin{multline}\label{kin-s-00}
\left( \frac{d}{dt} \right)_{ij} \overline{s_i^0s_j^0 s_{I'}^{P'}} = 2W_{ij} \overline{n_{j\uparrow} n_{j\downarrow} (1 - n_{i\uparrow})(1 - n_{i\downarrow}) s_{I'}^{P'} } - \\
 \frac{W_{ji}}{2} \overline{(s_i^0 s_j^0 - \sum_{\alpha} s_i^{\alpha}s_j^{\alpha} )s_{I'}^{P'}},
\end{multline}
\begin{equation} \label{kin-s-0a}
\left( \frac{d}{dt} \right)_{ij} \overline{s_i^0s_j^\alpha s_{I'}^{P'}} = \frac{W_{ji}}{2}
\overline{(s_i^{\alpha}s_j^0 - s_i^{0}s_j^\alpha) s_{I'}^{P'}},
\end{equation}
\begin{equation}\label{kin-s-a0}
\left( \frac{d}{dt} \right)_{ij} \overline{s_i^\alpha s_j^0 s_{I'}^{P'}} = \frac{W_{ji}}{2}
\overline{(s_i^{0}s_j^\alpha - s_i^{\alpha}s_j^0) s_{I'}^{P'}},
\end{equation}
\begin{multline}\label{kin-s-ab}
\left( \frac{d}{dt} \right)_{ij} \overline{s_i^\alpha s_j^\beta s_{I'}^{P'}} = \\
 -2W_{ij} \delta_{\alpha\beta} \overline{n_{j\uparrow} n_{j\downarrow} (1 - n_{i\uparrow})(1 - n_{i\downarrow}) s_{I'}^{P'} } + \\
 \frac{W_{ji}}{2} \delta_{\alpha\beta} \overline{(s_i^0 s_j^0 - \sum_{\alpha} s_i^{\gamma}s_j^{\gamma} )s_{I'}^{P'}} - \\
 \frac{W_{ji}}{2} \overline{(s_i^{\alpha}s_j^{\beta} - s_i^{\beta} s_j^{\alpha})s_{I'}^{P'}}
\end{multline}
Here Greek indexes $\alpha$, $\beta$ and $\gamma$ stand for spin polarizations along Cartesian axes.

The physical meaning of Eqs.~(\ref{kin-s-00}-\ref{kin-s-ab}) is as follows. Eq.~(\ref{kin-s-00}) and the first two terms in r.h.s. of Eq.~(\ref{kin-s-ab}) show that the hop from site $i$ to site $j$ is possible only when both sites are single-occupied and electron spins on these sites are in singlet state. It decreases the probability of single occupation of both sites. The backward hop from $j$ to $i$ leads to the single occupation of sites with electrons in singlet state. The third term in Eq.~(\ref{kin-s-ab}) leads to relaxation of the antisymmetric combinations $\overline{s_i^{\alpha}s_j^{\beta} - s_i^{\beta} s_j^{\alpha}}$. It can be shown that these combinations are related to a coherent combination of singlet and triplet states of the two electrons. When the electrons are either in the singlet or in a triplet state $\overline{s_i^{\alpha}s_j^{\beta} - s_i^{\beta} s_j^{\alpha}} =0$. However, when their state is a coherent combination of singlet and triplet $\overline{s_i^{\alpha}s_j^{\beta} - s_i^{\beta} s_j^{\alpha}}  \ne 0$.  Even if due to some reason the probabilities of singlet and triplet states are conserved, their coherent combinations relax because the hopping $i\leftrightarrow j$  can occur in the singlet state and cannot occur in a triplet state. Similar term appears in spin dynamic of a double quantum dot \cite{smirnov}. Eqs.~(\ref{kin-s-0a},\ref{kin-s-a0}) show that spin transfer process in $AB$ pairs of sites is correlated with occupation numbers.

The known relations between $\overline{c}_I^P$ and $\overline{s}_I^P$ allow to obtain the expressions for $(d/dt)_{ij}\overline{c}_I^P$ similar to Eqs.~(\ref{kin-s-00}-\ref{kin-s-ab}). These expressions are not provided here because they are quite cumbersome but can be given in a much more compact form when the correlation potentials are introduced.

The term $(d/dt)_i \overline{c}_{I}^P$ is responsible for the internal spin dynamics. It is present only when the index $p$ corresponding to site $i$ is a spin index.
\begin{equation}
\left(\frac{d}{dt}\right)_i \overline{c}_{i,I'}^{\alpha,P'} = \epsilon_{\alpha\beta\gamma}\Omega_{i,\beta}\overline{c}_{i,I'}^{\gamma,P'} - \frac{1}{\tau_s} \overline{c}_{i,I'}^{\alpha,P'}
\end{equation}
Here $\Omega_{i,\beta}$ is the projection of spin rotation frequency vector on site $i$ to the axis $\beta$. $\tau_s$ is a phenomenological on-site spin relaxation time. $\epsilon_{\alpha\beta\gamma}$ is Levi-Civita symbol.

In the linear response regime it is useful to introduce effective correlation potentials $\overline{\varphi}_{I}^P$.
\begin{equation} \label{pot-def}
\varphi_i^0 = c_i^0 / (1-n_i^{(0)})n_i^{(0)}, \quad \varphi_i^{\alpha} = c_i^{\alpha}/(s_i^0)_{eq}.
\end{equation}
Here $(s_i^0)_{eq}$ is the equilibrium probability of single occupation of site $i$. It is equal to $n_i^{(0)}$ if site $i$ has type $A$ and to $1-n_i^{(0)}$ if site $i$ has type $B$. There is no averaging in the definition (\ref{pot-def}). $\varphi_i^p$ should be ensemble averaged in some combination to have the meaning of potential. For example
$\overline{\varphi}_{ijk}^{0xy} = \overline{\varphi_i^0\varphi_j^x\varphi_k^y}$ can be considered as a potential of correlation $\overline{c}_{ijk}^{0xy}$.

In these notations $(d/dt)_{ik}\overline{c}_{i,I'}^{p,P'}$ corresponds to a ``correlation flow'' $J_{ik;I'}^{p;P'}$ between correlations $\overline{c}_{i,I'}^{p,P'}$ and
$\overline{c}_{k,I'}^{p,P'}$:
\begin{equation}\label{cor-ik}
\left( \frac{d}{dt} \right)_{ik} \overline{c}_{i,I'}^{p,P'} = - \left( \frac{d}{dt} \right)_{ik} \overline{c}_{k,I'}^{p,P'} =
J_{ik;I'}^{p;P'}.
\end{equation}
When $I'$ is the empty set $J_{ik}^0$ is the particle flow from site $k$ to site $i$.
The flow $J_{ik;I'}^{p;P'}$ is expressed as follows
\begin{multline}\label{cor-cur}
J_{ik;I'}^{p;P'} = \Gamma_{ik} \Theta_{I'}^{P'} \left( \overline{\varphi}_{k,I'}^{p,P'} - \overline{\varphi}_{i,I'}^{p,P'}+
S_{ik,I'}^{p} \right.
 + \\ \left.
\Phi_p^{qq'}(ik) \overline{\varphi}_{ik,I'}^{qq',P'} \right).
\end{multline}
Here $\Gamma_{ik}$ is the average number of electrons that hops from site $k$ to site $i$ in unit time in equilibrium (\ref{Gamma}).  $\Theta_{I'}^{P'}$ is the coefficient related to sites of $I$ that do not participate in transition $i \leftrightarrow k$.
\begin{equation}
\Theta_{j,l,m,...}^{p_j,p_l,p_m,...} = \theta_j^{p_j} \cdot \theta_l^{p_l} \cdot \theta_m^{p_m}\cdot...
\end{equation}
\begin{equation}
\theta_j^0 = n_j^{(0)}\left(1-n_j^{(0)}\right), \quad \theta_j^{\alpha} = \left(s_j^0\right)_{eq}.
\end{equation}
$S_{ik,I'}^p$ is the source term related to the external electrical field $\bf E$. It is not equal to zero only when $I'$ is empty set and $p=0$. In this case $S_{ik,\emptyset}^0 = e{\bf E} {\bf r}_{ik}$ where $e$ is electron charge, ${\bf r}_{ik}$ is vector of distance between $i$ and $k$.

The term $\Phi_p^{qq'}(ik) \overline{\varphi}_{ik,I'}^{qq',P'}$ describes the effect of higher-order correlations to the flow of lower-order correlations. For different indexes $p$, $q$ and $q'$, the coefficients $\Phi_p^{qq'}(ik)$ are
\begin{equation}\label{Phi}
\begin{array}{l}
\Phi_0^{00}(ik) = n_k^{(0)}-n_i^{(0)}, \\
\Phi_0^{\alpha\alpha}(ik) = \tau_k - \tau_i, \\
\Phi_\alpha^{\alpha 0}(ik) = (1-\tau_i)n_k^{(0)} - \tau_i(1-n_k^{(0)}), \\
\Phi_\alpha^{0\alpha}(ik) = \tau_k(1-n_i^{(0)}) - (1-\tau_k)n_i^{(0)}.
\end{array}
\end{equation}
Here $\tau_i=0$ for $A$-type site $i$ and $\tau_i=1$ if site $i$ has type $B$. All the coefficients $\Phi_p^{qq'}$ not listed in (\ref{Phi}) are equal to zero. For example when all the three indexes $p$, $q$ and $q'$ are spin indexes $\Phi_p^{qq'} = \Phi_\alpha^{\beta \gamma} = 0$.

The term $(d/dt)_{ij}\overline{c}_{ij,I'}^{pq,P'}$ is closely related to the correlation flow between correlations $\overline{c}_{i,I'}^{q',P'}$ and $\overline{c}_{k,I'}^{q',P'}$:
\begin{multline}\label{cur_up}
\left(\frac{d}{dt}\right)_{ij}\overline{c}_{ij,I'}^{pq,P'} = - \Phi^{pq}_{q'}(ij) J_{ij;I'}^{q';P'} - \\
\Upsilon_{ij}^{pq} \Theta_{ij,I'}^{pq,P'} \left(  \overline{\varphi}_{ij,I'}^{pq,P'} - \overline{\varphi}_{ij,I'}^{qp,P'} \right)
\end{multline}
Note that the same coefficients $\Phi_p^{qq'}$ enter the equations (\ref{cor-cur}) and (\ref{cur_up}).
The second term in r.h.s. of Eq.~(\ref{cur_up}) is related to the relaxation of coherent singlet-triplet combinations. It contains the coefficient $\Upsilon_{ij}^{pq}$ that is equal to unity when sites $i$ and $j$ have different types and $p$ and $q$ are spin indexes. Otherwise, $\Upsilon_{ij}^{pq} = 0$.

The term $(d/dt)_i \overline{c}_{i,I'}^{\alpha,P'}$ related to the spin rotation and relaxation should also be expressed in terms of potentials
\begin{equation}\label{cur_int}
\left(\frac{d}{dt}\right)_i \overline{c}_{i,I'}^{\alpha,P'} = \Theta_{i,I'}^{\alpha,P'} \left( \epsilon_{\alpha\beta\gamma} \Omega_{i,\beta} \overline{\varphi}_{i,I'}^{\gamma,P'} - \frac{\overline{\varphi}_{i,I'}^{\alpha,P'}}{\tau_s} \right).
\end{equation}

In a stationary system the derivatives $d \overline{c}_I^P/dt$ are equal to zero. Therefore the equations (\ref{kin-gen-1}),
(\ref{cor-ik}),(\ref{cor-cur}), (\ref{cur_up}) and (\ref{cur_int}) compose a closed system of linear equations for the correlations potentials. It incudes all the charge and spin correlations. The total number of these correlations is extremely large, $4^N$ where $N$ is the number of sites. However, one can hope that correlations between sites at very large distances and the correlations of very high order are not relevant for the electron transport and can be neglected. Actually, to treat reasonably large systems some of the correlations should be neglected to make the system of equations solvable. The idea of CKE approach is to write the equations in general form relevant for arbitrary correlations and make the cutoff at the ``final step'' taking into account the structure of considered system (that defines the correlations that are really relevant) and the possibility to numerically solve the system of equations of the desired size. When some correlations are neglected in this way, the potentials $\overline{\varphi}_I^P$ of these  correlations are considered to be equal to zero in all the equations.

\section{Magnetoresistance due to the relaxation of spin correlations}
\label{s:MR}

In this section, the discussed approach to the theory of hopping transport with spin correlations is applied to the bipolaron mechanism of OMAR. As it was discussed in the previous section, it is necessary to cut the system of kinetic equations at some point. The most simple cutoff is the model when only the correlations in close pairs of sites are considered. In this case, it is possible to reduce the problem to a network of modified Miller-Abrahams resistors. This approach was used in \cite{Aleiner2, AVS-cor} with the semi-classical model of spins up and down.  However, in the present study, the quantum nature of spin correlations is taken into account and the expressions for resistors are different from \cite{AVS-cor}. This model is discussed in Sec.~\ref{s:MR-an}. It allows the analytical solution in the limiting cases of fast and slow hopping.

The long-range and high-order correlations can be taken into account with the numerical solution of kinetic equations. Such solutions are provided in Sec.~\ref{s:MR-num} and are compared with analytical results.

\subsection{Modified resistor model}
\label{s:MR-an}

When the long-range spin correlations are neglected the rate equation for spin correlations $\overline{c}_{ij}^{\alpha\beta}$ includes only the spin generation due to the electron flow $J_{ij}$ and the internal spin dynamics. The effect of other sites is reduced to spin relaxation. When the correlation $i-j$ is considered in this model the spins on other sites are assumed to be in thermal equilibrium. The transition of correlation $i-j$ to other sites  can be formally included into internal spin dynamics as additional source of relaxation.
\begin{equation}
\frac{d}{dt} \overline{c}_{ij}^{\alpha\beta} = -R_{\alpha\beta;\alpha'\beta'}^{(ij)}\overline{c}_{ij}^{\alpha'\beta'} +
\delta_{\alpha\beta}(\tau_i - \tau_j)J_{ij}^{0}.
\end{equation}
Here ${R}_{\alpha\beta;\alpha'\beta'}^{(ij)}$ is a matrix that describes the dynamics and relaxation of correlations.
\begin{multline}\label{R1}
{ R}_{\alpha\beta;\alpha'\beta'}^{(ij)} =  \gamma_{ij} \delta_{\alpha\alpha'}\delta_{\beta\beta'} - \epsilon_{\alpha\gamma\alpha'}\delta_{\beta\beta'} \Omega_{i,\gamma} - \\ \epsilon_{\beta\gamma\beta'} \delta_{\alpha\alpha'}\Omega_{j,\gamma} + \frac{\Gamma_{ij}}{(s_i^0)_{eq} (s_j^0)_{eq}}(\delta_{\alpha\alpha'}\delta_{\beta\beta'} - \delta_{\alpha\beta'}\delta_{\beta\alpha'}).
\end{multline}
$\gamma_{ij}$ is the effective rate of relaxation of spin correlations either due to phenomenological on-site spin relaxation mechanism or due to electron transition to or from other sites.
\begin{equation}
\gamma_{ij} = \frac{1}{\tau_s} + \sum_{k\ne i,j}\left( T_{ki}^{(\alpha)} + T_{kj}^{(\alpha)} \right).
\end{equation}
Note that spin transition rates $T_{ki}^{\alpha}$ do not depend on value of spin index $\alpha = x,y,z$. The index is kept only to show that it is a spin index, not the charge index 0. It is assumed in the present section that $i-j$ is the resistor that controls the resistivity of some mesoscopic part of the sample. It happens when the rate of hopping inside the pair $i-j$ is slow compared to the hopping between this pair and other sites. It allows to assume that $\Gamma_{ij} \ll \gamma_{ij}$. Therefore $\Gamma_{ij}$ and the last term in r.h.s. part of Eq.~(\ref{R1}) is neglected in the present section.

It is possible to give a closed expression for $J_{ij}^0$ with account to short-range pair correlations in terms of inverse matrix $({R}^{(ij)})^{-1}$.
\begin{equation}
J_{ij}^0 = \frac{\varphi_j^0 - \varphi_i^0}{\Gamma_{ij}^{-1} + {\cal F}_s(ij) + {\cal F}_c(ij)},
\end{equation}
\begin{equation}\label{Fs}
{\cal F}_s(ij) = \frac{ \delta_{\alpha\beta}\delta_{\alpha'\beta'} \left({R}^{(ij)}\right)^{-1}_{\alpha\beta;\alpha'\beta'} (\tau_i -\tau_j)^2}{(s_{i}^0)_{eq} (s_j^0)_{eq}},
\end{equation}
\begin{equation}
{\cal F}_c (ij) = \frac{\left(n_i^{(0)} - n_j^{(0)}\right)^2 \left( \sum_{k \ne i,j} T_{ki}^{(0)} + T_{kj}^{(0)} \right)^{-1}}{n_i^{(0)}n_j^{(0)} (1-n_i^{(0)}) (1-n_j^{(0)}) }.
\end{equation}
Here $\Gamma_{ij}^{-1}$ corresponds to ordinary Miller-Abrahams resistance. ${\cal F}_s$ and ${\cal F}_c$ describe the additional resistance that appear due to spin and charge correlations correspondingly.

The effect of the external magnetic field on the conductivity is incorporated in ${\cal F}_s$. As shown in eqs.~(\ref{R1},\ref{Fs}) it depends on the vectors of on-site rotation frequencies $\mathbf{\Omega}_i$ and $\mathbf{\Omega}_j$ that are proportional to the sum of hyperfine field and applied external magnetic field  $\mathbf{\Omega}_i = \mu_b g ({\bf H} + {\bf H}_{\rm hf}^{(i)})/\hbar$. However, even if the system is composed from only one resistor the averaging over $\mathbf{\Omega}_i$ and $\mathbf{\Omega}_j$ is required. The hyperfine fields are slowly changed due to the nuclear spin dynamics. Although this dynamics is considered to be slow compared to electron hops and electron spin rotation, it is usually fast compared to the current measurement procedures. Therefore the final expression for dc resistivity should be averaged over hyperfine fields.

I assume that hyperfine fields have normal distribution
\begin{equation}
p(H_{{\rm hf},\alpha}^{(i)}) = \frac{1}{\sqrt{2\pi}H_{\rm hf}} \exp \left[ - \frac{\left(H_{{\rm hf},\alpha}^{(i)}\right)^2}{2H_{\rm hf}^2} \right].
\end{equation}
Here $H_{\rm hf}$ is the typical value of hyperfine fields. The different components of the hyperfine field on a given site and the fields on different sites are considered not to be correlated.

The spin correlation part of resistance is related to the reverse relaxation function ${\cal R}(\gamma_{ij}, H, H_{\rm hf})$: ${\cal F}_s = (\tau_i-\tau_j)^2 {\cal R}/(s^{0}_i)_{eq}(s^{0}_j)_{eq}$.
\begin{equation}
{\cal R} = \left\langle \delta_{\alpha\beta}\delta_{\alpha'\beta'} (R^{(ij)})^{-1}_{\alpha\beta;\alpha'\beta'} \right\rangle_{\rm hf}.
\end{equation}
Here $\langle ... \rangle_{\rm hf}$ means the averaging over the hyperfine fields. ${\cal R}$ can be thought of as the time of relaxation of probabilities for the two spins to be in singlet or triplet state. In the general case ${\cal R}(\gamma_{ij}, H, H_{\rm hf})$ can be found numerically. However, it is possible to find it analytically in the limiting cases of slow and fast hops.

\begin{figure*}[htbp]
    \centering
        \includegraphics[width=0.8\textwidth]{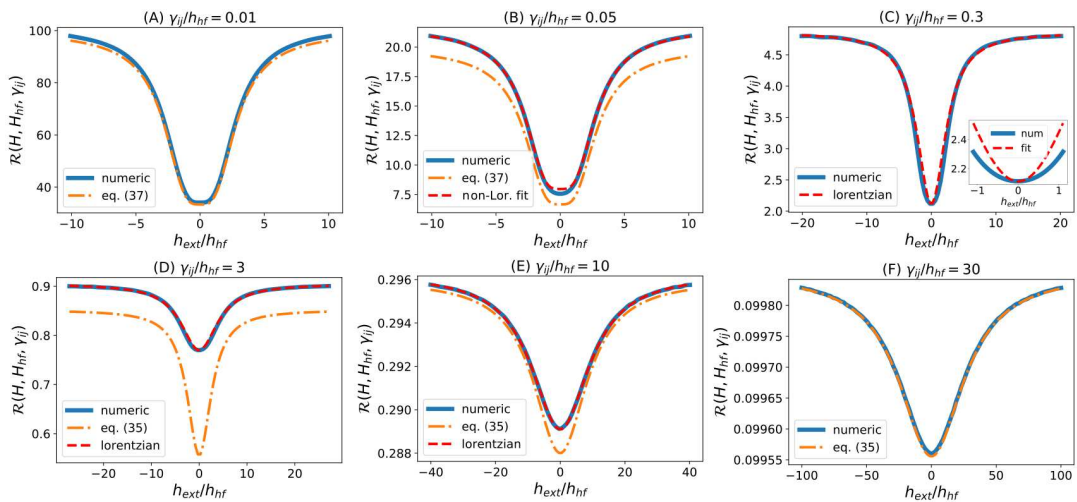}
        \caption{Reverse relaxation function ${\cal R}(h_{\rm ext}/h_{\rm hf})$ corresponding to different relations $\gamma_{ij}/h_{\rm hf}$. Solid blue curve in all the figures is numeric calculation of ${\cal R}$. Yellow dash-dot curve in Fig. (A) and (B) is approximation with Eq.~(\ref{R4}). Red dashed curve in Fig. (B) is the non-Lorentz fit described in text. Red dashed curve on Fig.(C)-(D) is the Lorentzian fit. Yellow dash-dot curve in Fig. (D)-(F) is approximation with Eq.~(\ref{R2}). Inset in Fig.~(C) shows the curves at small magnetic fields.  }
    \label{fig:revrel}
\end{figure*}

In the limit of  fast hops the rate of relaxation $\gamma_{ij}$ due to electron transition to other sites is fast compared to the typical rate of rotation in hyperfine fields $h_{\rm hf} = \mu_b g H_{\rm hf}/\hbar$. The rate of rotation in the external magnetic field $h_{\rm ext} = \mu_b g H/\hbar$ is arbitrary. In this case relaxation matrix $R^{(ij)}$ can be divided into $R^{(ij)}_1$ related to hopping and rotation in the external magnetic field $R^{(ij)}_1 = - \gamma_{ij} \delta_{\alpha\alpha'}\delta_{\beta\beta'} + \mu_b g H(\epsilon_{\alpha z \alpha'}\delta_{\beta\beta'}  + \epsilon_{\beta z \beta'} \delta_{\alpha\alpha'})/\hbar$ and $R^{(ij)}_2$ related to the rotation in hyperfine fields. The first of this matrices $R^{(ij)}_1$ can be inverted analytically. The second one can be considered as a small perturbation. The total inverse relaxation matrix averaged over hyperfine fields can be approximately expressed as
\begin{multline}\label{R1a}
\left\langle \left(R^{(ij)}\right)^{-1} \right\rangle_{\rm hf}= \left(R^{(ij)}_1\right)^{-1} + \\ \left\langle \left(R^{(ij)}_1\right)^{-1} R^{(ij)}_2  \left(R^{(ij)}_1\right)^{-1}  R^{(ij)}_2 \left(R^{(ij)}_1\right)^{-1}
\right\rangle_{\rm hf}.
\end{multline}
In principle the expression for  $(R^{(ij)})^{-1}$ also includes the first order term $(R^{(ij)}_1)^{-1} R^{(ij)}_2  (R^{(ij)}_1)^{-1}$, however, it becomes equal to zero after the averaging over hyperfine fields.

With straightforward calculations, Eq.~(\ref{R1a}) leads to explicit expression for the function ${\cal R}$
\begin{equation}\label{R2}
{\cal R} = \frac{1}{\gamma_{ij}} \left[3 - 4 \frac{h_{\rm hf}^2}{\gamma_{ij}^2} \left(1 + \frac{2}{1 + h_{\rm ext}^2/\gamma_{ij}^2}\right) \right].
\end{equation}
The dependence of resistance on the magnetic field corresponding to eq.~(\ref{R2}) is described by Lorentz function. Note that its width is controlled not by the relation of external magnetic and hyperfine fields $h_{\rm ext}/h_{\rm hf}$ but by the relation of hopping rate ant the rate of rotation in the external field $h_{\rm ext}/\gamma_{ij}$. The magnetoresistance is relatively weak due to the small prefactor $h_{\rm hf}^2/\gamma_{ij}^2$.

The opposite limit is the situation of slow hopping, $\gamma_{ij} \ll h_{\rm hf}$. In this case it is possible to use the random phase approximation. It is assumed that component of electron spin on site $i$ normal to the on-site effective magnetic field ${\bf H} + {\bf H}_{\rm hf}^{(i)}$ relaxes very fast due to the rotation around this field. However, the component along ${\bf H} + {\bf H}_{\rm hf}^{(i)}$ is conserved until the electron is transferred to some other site. In this case ${\cal R}$ is proportional to the averaged squared cosinus of the angle between on-site fields
\begin{equation}\label{R3}
{\cal R} = \left\langle \cos^2 \left({\bf H}^{(i)}, {\bf H}^{(j)}\right) \right\rangle_{\rm hf}/\gamma_{ij}.
\end{equation}
Here ${\bf H}^{(i)} = {\bf H} + {\bf H}_{\rm hf}^{(i)}$. The averaging in (\ref{R3}) can be done analytically in terms of special functions
\begin{multline}\label{R4}
{\cal R} = \frac{1}{3\gamma_{ij}} + \frac{12}{\gamma_{ij}H^6} \times \\ \left[
\frac{\sqrt{2}}{6}H(H^2-3H^2_{\rm hf}) + H_{\rm hf}^3 D_+\left( \frac{H}{\sqrt{2}H_{\rm hf}}\right)
\right]^2.
\end{multline}
Here $D_+(x)$ is the Dawson function $D_+(x) = e^{-x^2} \int_0^x e^{t^2} dt$. Eq.~(\ref{R4}) shows that in slow hopping limit the dependance of resistance on magnetic field has non-Lorentz shape. It saturates when applied magnetic field is much larger than hyperfine fields, while $\gamma_{ij}$ controls its overall strength.

In Fig.~\ref{fig:revrel} I compare the approximate expressions (\ref{R2}) and (\ref{R4}) with the function ${\cal R}$ calculated numerically. It can be seen that Eq.~(\ref{R2}) can quantitatively describe ${\cal R}$ only for quite large values of hopping rate $\gamma_{ij} \gtrsim 30h_{\rm hf}$. However, the dependence of ${\cal R}$ on the applied magnetic field can be described by the Lorentzian $R = A + B/(h_{\rm ext}^2 + \widetilde{\gamma}^2 )$ for significantly smaller $\gamma_{ij} \gtrsim 3h_{\rm hf}$. Here $A$, $B$ and $\widetilde{\gamma}$ are fitting parameters. At smaller hopping rates $\gamma_{ij} \lesssim 0.3 h_{\rm hf}$ the function ${\cal R}$ becomes non-Loretzian. It is most clearly seen when comparing the numeric results for ${\cal R}$    with its Lorentzian fit at small magnetic fields as shown on the inset in Fig.~\ref{fig:revrel}(C).

For small hopping rates $\gamma_{ij} \lesssim 0.05h_{\rm hf}$ the non-Lorentzian fit related to Eq.~(\ref{R4}) becomes relevant. It is the fit ${\cal R} = A+B\widetilde{\cal R}(H,H_{\rm hf},\gamma_{ij})$ where $\widetilde{\cal R}$ is described with Eq.~(\ref{R4}). For $\gamma_{ij} \lesssim 0.01 h_{\rm hf}$  Eq.~(\ref{R4}) can describe the reverse relaxation function quantitatively.

It is interesting to compare the results of the present section with results of \cite{AVS-cor} where the correlations in close pairs of sites were considered with semi-classical model of spin. In \cite{AVS-cor} OMAR was related to spin relaxation time $\tau_s(H)$ that had a phenomenological dependence on the applied magnetic field. The approach of the present section can describe quantum spin correlations with the same relaxation time. When the spin relaxation is reduced to a single time $\tau_s(H)$ the reverse relaxation function can be found analytically and the spin part of resistance is equal to ${\cal F}_s = 3(\tau_i-\tau_j)^2\tau_s(H)/[(1+\gamma_{ij}\tau_s(H))(s_i^0)_{eq} (s_j^0)_{eq} ]$. It is exactly 3 times larger than the correction  to Miller-Abrahams resistor due to spin correlations obtained in \cite{AVS-cor}. Note that ${\cal F}_c$ quantitatively agree with the correction to Miller-Abrahams resistor due to charge correlations obtained in \cite{AVS-cor}. The difference between the results is related to the quantum nature of spin correlations taken into account in the present paper. It appears that semi-classical model of spin cannot quantitatively describe OMAR even if the spin relaxation is reduced to a single time $\tau_s(H)$.

\subsection{Numerical results}
\label{s:MR-num}

\begin{figure}[htbp]
    \centering
        \includegraphics[width=0.37\textwidth]{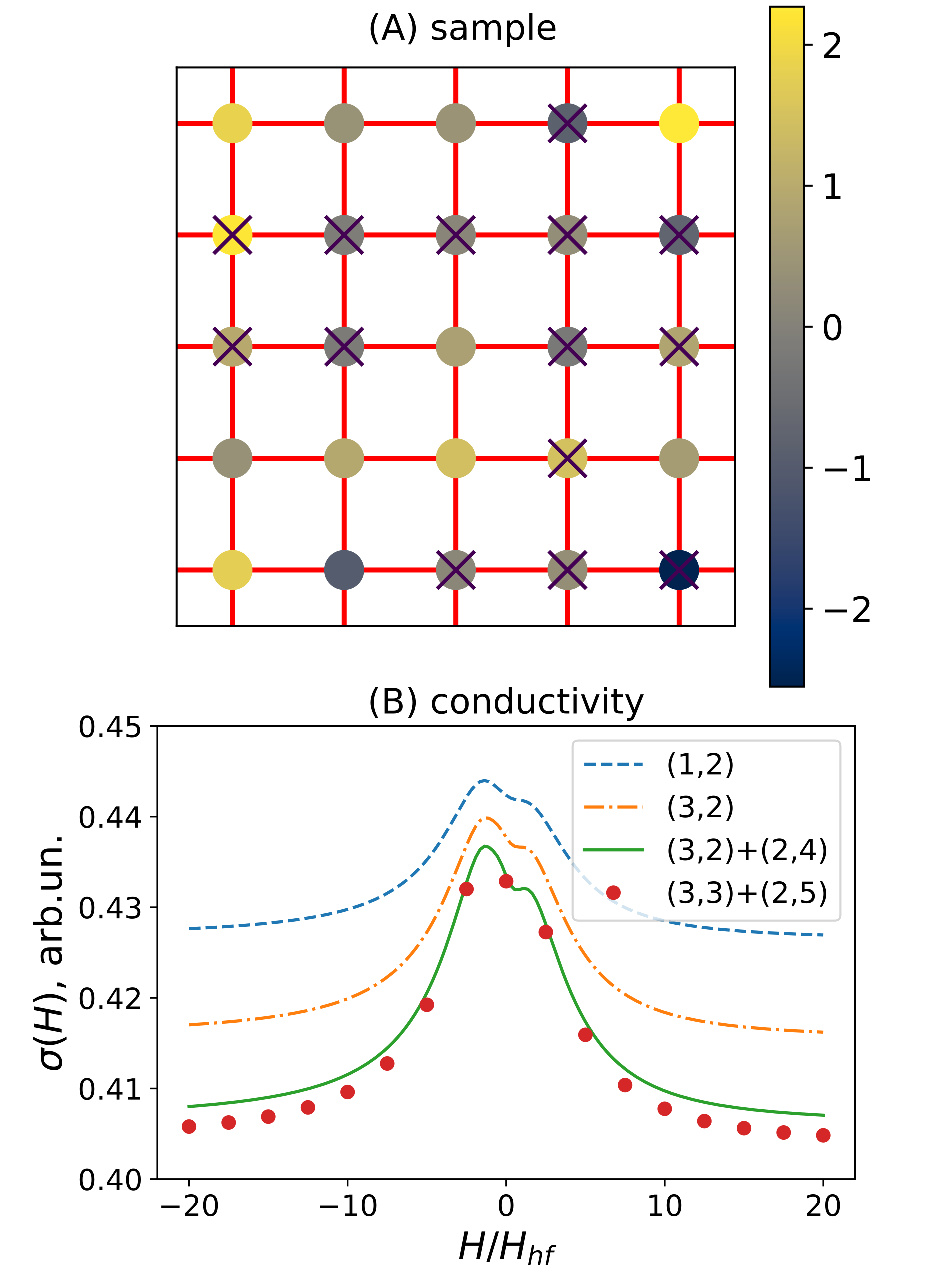}
        \caption{The dependence of conductivity on the applied magnetic field in mesoscopic numeric sample. (A) the structure of the sample. The color bar stands for site energies. (B) conductivity calculated with different approximations as described in the text. }
    \label{fig:smallsample}
\end{figure}

This section includes the results of the numerical solution of correlation kinetic equations for several disordered systems. The calculation is made with some of the long-range and high-order correlations taken into account. The results are compared with the model described in Sec.~\ref{s:MR-an} to show when the theory that includes only the correlations in close pairs of sites is applicable and when it is not.

I start from a single numerical sample consisting of 25 sites. The sample is shown in Fig.~\ref{fig:smallsample}(A). The sites are placed on a square lattice with the same overlap integrals $t_{ij}$ between neighbor sites. The site energies are selected independently with normal distribution with the standard deviation $\Delta E$. The temperature is $T = \Delta E$. Some sites are randomly selected to have type B. They are marked with crosses in Fig.~\ref{fig:smallsample}(A). Other sites have type A. Each site is ascribed with a random hyperfine field. The typical rate of rotation in  random fields is equal to the pre-exponential term in the hopping rate between neighbors $h_{\rm hf} = w_0$. Here I define the pre-exponential term in the hopping rates as $w_0 = W_0|t_{ij}|^2$. The periodic boundary conditions are applied.

\begin{figure*}[htbp]
    \centering
        \includegraphics[width=1.0\textwidth]{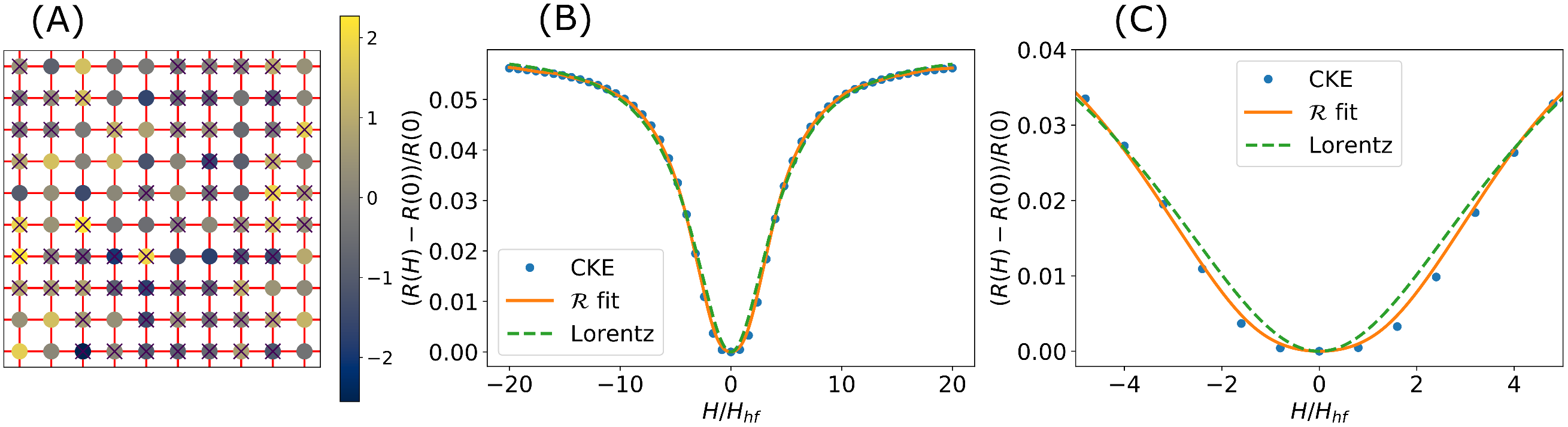}
        \caption{Results of CKE solution for random $10\times 10$ numerical samples. (A) the structure of a single sample. (B) and (C) calculated magnetoresistance in different scales averaged over 10 random samples. }
    \label{fig:normalsample}
\end{figure*}

In Fig.~\ref{fig:smallsample}(B) the calculated conductivity of this sample is presented. The conductivity is calculated in four different approximations. The blue dashed curve corresponds to $(1,2)$ approximation where pair correlations between sites $i$ and $j$ are included in the theory when the distance between sites $i$ and $j$ along the lattice bonds is 1. It is the approximation used in Sec.~\ref{s:MR-an}. In general, the notation $(p,q)$ stands for the approximation when correlations up to the order $q$ are considered provided that the distance between sites in the correlations along lattice bonds is no longer than $p$. The yellow dash-dot curve corresponds to $(3,2)$ approximation when most pair correlations are taken into account. Green solid curve corresponds to joint $(3,2)$ and $(2,4)$ approximation when most of pair correlations and correlations up to the 4-th order in close complexes of sites are considered. The red dots stand for joint $(3,3)$ and $(2,5)$ approximation where most of the correlations of the third order and the correlations up to 5-th order in close complexes are included in the theory.

The introduction of new correlations into the theory decreases the calculated conductivity and increases its calculated dependence on the applied magnetic field. However, the difference between $(3,2)+(2,4)$ and $(3,3)+(2,5)$ approximations is rather small and one can hope that $(3,2)+(2,4)$ approximation adequately describes the system. In the following analysis of larger numerical samples, the correlations of order $q>2$ will be considered only for the distance between sites $p \le 2$. The pair correlations will be considered at slightly larger distances. Unfortunately, the number of spin correlations grows extremely fast with the correlation order and it was technically impossible to go beyond the $(3,3)+(2,5)$ approximation even for the quite small $5 \times 5$ sample.

Note that the magnetic field dependence of conductivity shown in Fig.~\ref{fig:smallsample}(B) is not symmetric with respect to the inversion of the sign of the magnetic field. It is related to the mesoscopic nature of the considered numerical sample. There is a finite number of sites and each site is ascribed with the well-defined on-site hyperfine field. It breaks the time-reversal symmetry of the calculated system. In real samples (even in mesoscopic ones) the hyperfine fields slowly change in time. It restores the time-reversal symmetry. Therefore the dependence $\sigma(H)$ for real samples is symmetric even if they are mesoscopic.

In Fig.~\ref{fig:normalsample} I show results for larger $10 \times 10$ samples. The results are averaged over 30 disorder configurations. The properties of the numerical samples (except their size) are the same as in Fig.~~\ref{fig:smallsample}. Fig.~\ref{fig:normalsample}(A) shows the structure of one of these samples. Fig.~\ref{fig:normalsample}(B) and (C) show the magnetoresistance $(R(H) - R(0))/R(0)$ in different scales. Here $R(H)$ is the sample resistance in the external magnetic field $H$. The blue points stand for the magnetoresistance calculated with the numeric solution of CKE. The pair correlations at the distance no longer than $4$ and 4-th order correlations with the distance $2$ were taken into account ($(4,2)+(2,4)$ approximation). Green dashed curve is the fit with Lorentzian. Yellow solid curve is the fit with function ${\cal R}$ described in Sec.~\ref{s:MR-an}. It means that the expression
\begin{equation}\label{Rfit}
\frac{R(H)-R(0)}{R(0)} = A \times \frac{{\cal R}(\widetilde{\gamma}, H, H_{\rm hf}) -  {\cal R}(\widetilde{\gamma}, 0, H_{\rm hf}) }{{\cal R}(\widetilde{\gamma}, 0, H_{\rm hf})}
\end{equation}
was used for fitting. Here $A$ and $\widetilde{\gamma}$ are the fitting parameters. $\widetilde{\gamma}$ can be considered as the effective rate for a correlation to leave the pair of sites where it appeared. The results shown in Fig.~\ref{fig:normalsample}(B) correspond to $\widetilde{\gamma} \approx 2.3 h_{\rm hf}$. $A$ is the general amplitude of magnetoresistance, it describes the relative part of sample resistance that is related to spin correlations.

The shape of resistance dependence on the magnetic field significantly deviates from Lorentzian. It can be easily seen in Fig.~\ref{fig:normalsample}(C) where the results for small magnetic fields are shown on a close scale. The fitting (\ref{Rfit}) has better agreement with numerical simulation.

\begin{figure}[htbp]
    \centering
        \includegraphics[width=0.5\textwidth]{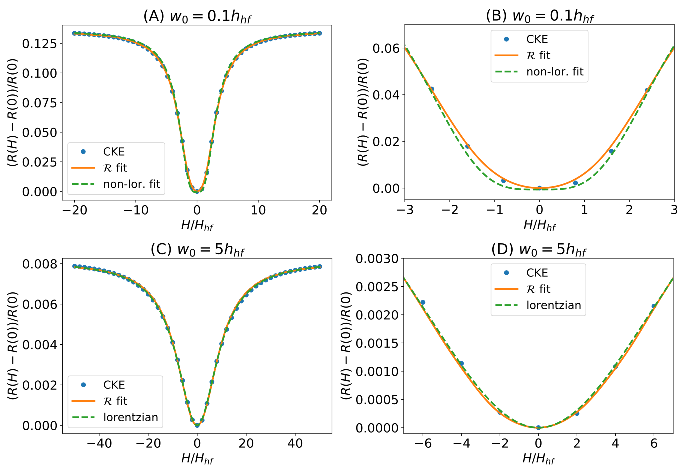}
        \caption{Magnetoresistance in numerical samples with slow $w_0 = 0.1h_{\rm hf}$ and fast $w_0 = 5h_{\rm hf}$ hopping rates. Blue dots on all subplots correspond to results of numerical solution of CKE. Yellow solid curves correspond to Eq.~(\ref{Rfit}). Green dashed curves in Figs (A) and (B) is the fit with Eq.~(\ref{Rfit}) where function ${\cal R}$ is described by Eq.~(\ref{R4}) and fitting hyperfine fields are artificially increased $h_{\rm hf} \rightarrow 1.08h_{\rm hf}$. Green dashed curve in Figs (C) and (D) is the fit with Lorentzian.  }
    \label{fig:difw0}
\end{figure}

In Fig.~\ref{fig:difw0} the similar analysis is provided for similar numeric samples with different rates of hopping between neighbors $w_0 = 0.1h_{\rm hf}$  and $w_0 = 5h_{\rm hf}$. Other characteristics of the samples are the same as in the previous numeric experiment including the averaging over 30 disorder configurations. Fig.~\ref{fig:difw0} (A) and (B) shows the results for the samples were the hopping is slow, $w_0 = 0.1 h_{\rm hf}$. Blue points correspond to the numeric solution of CKE. Yellow solid curve corresponds to the fit with Eq.~(\ref{Rfit}), the fitting parameter $\widetilde{\gamma}$ was $\widetilde{\gamma} = 1.1h_{\rm hf}$. Green dashed curve corresponds to fit with Eq.~(\ref{Rfit}), where function ${\cal R}$ is described by Eq.~(\ref{R4}) (that is valid in the limit $\widetilde{\gamma} \ll h_{\rm hf}$) but the strength of hyperfine fields was artificially increased $h_{\rm hf} \rightarrow 1.08 h_{\rm hf}$ to achieve better fit with numerical results.

\begin{figure*}[htbp]
    \centering
        \includegraphics[width=0.85\textwidth]{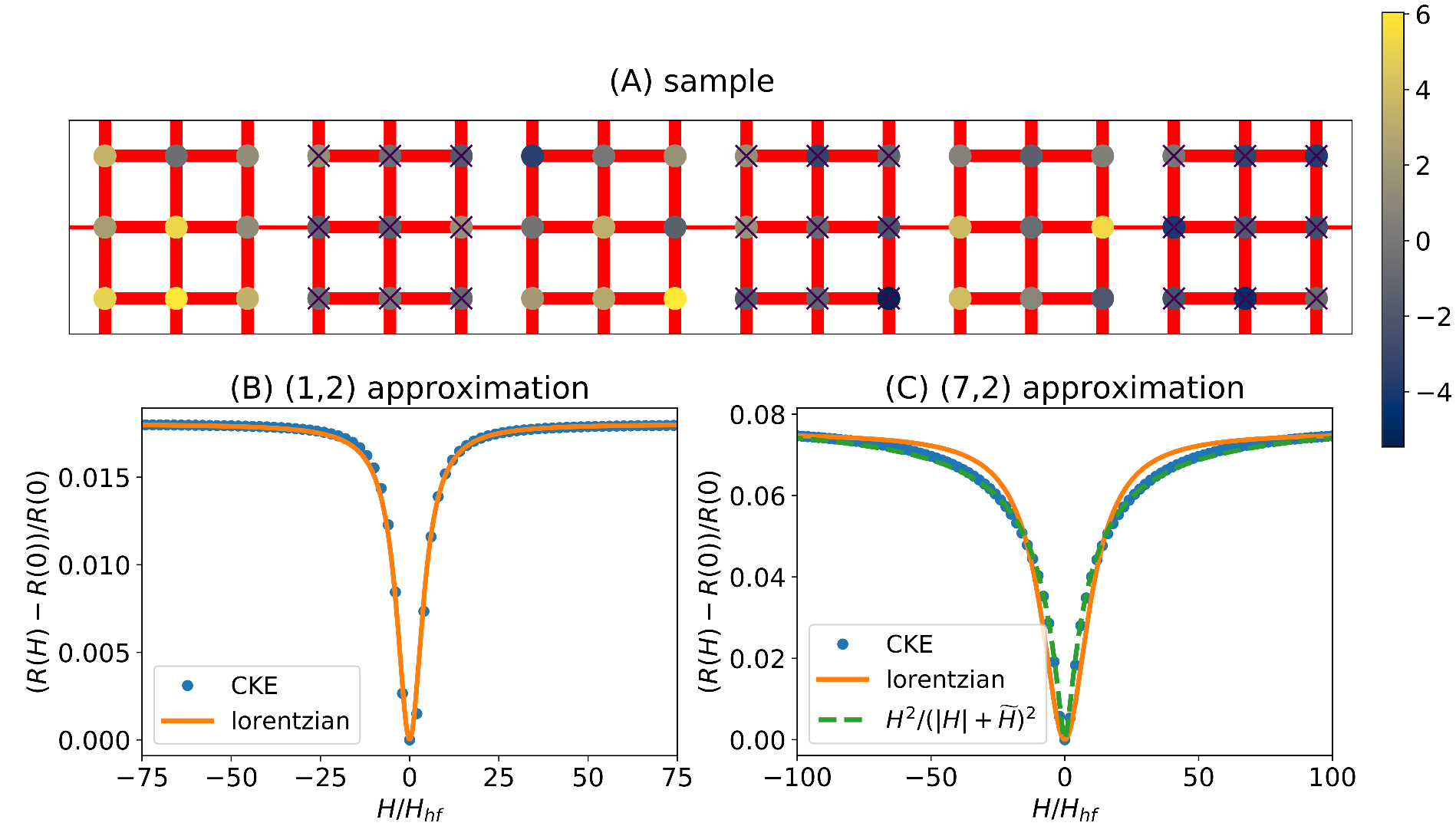}
        \caption{Results for the numeric samples constructed with $3\times3$ blocs. (A) the structure of a numeric sample. (B) magnetoresistance in $(1,2)$ approximation. (C) magnetoresistance in (7,2) approximation.  }
    \label{fig:specsample}
\end{figure*}

Note that the model from Sec.~\ref{s:MR-an} does not take into account long-range and high-order correlations. Therefore, the possibility of the description of numerical results with this model could not be taken for granted. However, for the considered numerical samples, this description is possible and the effect of the simplifications made  in Sec.~\ref{s:MR-an} is reduced to small modification of $h_{\rm hf}$ and to some changes in the amplitude of magnetoresistance (described with fitting parameter $A$).

In Fig.~\ref{fig:difw0} (C) and (D) the results for numeric samples with fast hopping $w_0 = 5 h_{\rm hf}$ are shown. The results of the simulation (blue dots) agree with fitting with Eq.~(\ref{Rfit}) (yellow solid curve), where the value of fitting parameter $\widetilde{\gamma}$ is $8.7 h_{\rm hf}$. At this $\widetilde{\gamma}$ the function ${\cal R}$ can be described by Lorentzian, as shown with green dashed curve in Fig.~\ref{fig:difw0} (C) and (D).

The provided numeric results show that in some cases the theory from Sec.~\ref{s:MR-an} can be used as a toy model for the understanding more complex situations when long-range and high-order correlations are required to quantitatively calculate the magnetoresistance. However, not all the lineshapes that appear in numeric experiments can be described with this toy model. Let us consider the numeric sample shown in Fig.~\ref{fig:specsample}. It is constructed from $3\times3$ blocks, inside a block the sites have the same type and the hopping between them is fast $w_{0} = 3h_{\rm hf}$. The blocks are connected with links with slow hopping $w_{0} = 0.3 h_{\rm hf}$. The energies of sites are random with normal distribution with the standard deviation $\Delta E = T$. The average energy of sites in $A$-type block is $1.5T$ and in $B$-type block it is $-1.5T$. The idea behind this sample is as follows. The conductivity of the sample is controlled by the process of generation and recombination of electron-hole pairs in $AB$ pairs of sites. The structure of the sample ensures that the generated electron and hole will stay near the pair of sites were they are generated for quite a long time. However, they are trapped not on a single site but on a cluster of nine sites, therefore their spins not only rotate around local hyperfine fields but also relax due to hops between sites with different hyperfine fields.  The positive average energy of $A$-type sites and the negative one of $B$-type sites ensures that $A$-type clusters contain a small number of electrons and $B$-type clusters contain a small number of holes. Physically this situation can correspond to small polymer molecules where the hopping between monomers of a single polymer is fast while the hops between different molecules are slow. Note that for the relaxation of spin correlation in the discussed electron-hole pair all the $18$ hyperfine fields in two neighbor blocks are relevant. It cannot be captured with the toy model from Sec.~\ref{s:MR-an}.

The structure of the described sample is shown in Fig.~\ref{fig:specsample} (A).  Fig.~\ref{fig:specsample} (B) shows the magnetoresistance of such samples calculated in the $(1,2)$ approximation that corresponds to the model of modified Miller-Abrahams resistor. In Fig.~\ref{fig:specsample}(C) the magnetoresistance calculated in $(7,2)$ approximation is shown. In both of the cases, the magnetoresistance is averaged over $30$ random disorder configurations (i.e. the random hyperfine fields and site energies).

The magnetoresistance calculated in $(1,2)$ approximation is quite weak $\sim 2\%$ and its lineshape is Lorentzian. However, $(1,2)$ approximation is not adequate for the description of the magnetoresistance in these samples because the correlations can easily leave the initial sites but are trapped in the clusters. To take this trapping into account it is required to consider the correlations at the inter-site distance equal to $7$. When these correlations are taken into account the estimated magnetoresistance increases $\sim 4$ times and its lineshape becomes non-Lorentzian. It can be described with the expression
\begin{equation}\label{exp-fit}
\frac{R(H)-R(0)}{R(0)} \propto \frac{H^2}{(|H| + H_0)^2},
\end{equation}
where $H_0$ is a fitting parameter. The expression (\ref{exp-fit}) was used to describe the lineshape of OMAR in a number of experimental works \cite{OMAR,sheng, TDNguen, Shakya, Yan}.

I want to stress that the statistics of hyperfine fields is exactly the same in all the considered numerical samples. However, the obtained OMAR lineshapes are different including the two shapes most commonly obtained in experiments: $H^2/(H^2 + H_{0}^2)$  and $H^2/(|H| + H_0)^2$. What is different in the numerical samples is the statistics of electron hops. One can conclude that the shape of OMAR contains information about short-range transport in organic materials.

\section{discussion}
\label{s:dis}

Up to the recent time, the most known method for the calculation of hopping transport with the account to correlations of filling numbers was the numeric Monte-Carlo simulation. When only the charge correlations are important it leads to the correct description of transport provided that the time of simulation is enough to achieve the averaging. In \cite{CKE0} the Monte-Carlo simulation was used to prove that CKE approach also leads to correct results when a sufficient number of correlations are taken into account. Therefore the Monte-Carlo simulation was considered to be ``an arbiter'' for CKE approximations.

However, the situation is different when the spin correlations are relevant. In \cite{Bobbert} the Monte-Carlo simulation with the semi-classical description of electron spins in terms of ``spin up'' and ``down'' was used to show the possibility of bipolaron mechanism of OMAR. However, this simplified description cannot include spin rotation around hyperfine on-site fields. Naturally, ``up'' and ``down'' spins cannot rotate. Therefore the spin relaxation in \cite{Bobbert} was described by a single relaxation time $\tau_s$ with phenomenological dependence on the applied magnetic field.

Even when the spin relaxation is reduced to a single time $\tau_s$ the semi-classical description of spin does not lead to the correct qualitative estimate of the spin-correlation part of resistance. In \cite{AVS-cor} the semi-classical spin correlations were considered in the approximation when only pair correlations in close pairs of sites are taken into account. It leads to correlation corrections of Miller-Abrahams resistors. These corrections are compared with similar corrections due to quantum spin correlations in Sec.~\ref{s:MR-an}. The corrections due to charge correlations obtained in \cite{AVS-cor} and in the present study agree. However, the corrections due to spin correlations obtained in Sec.~\ref{s:MR-an} are exactly three times larger than the spin corrections to resistances in \cite{AVS-cor}. I believe that the reason for this difference is related to the quantum nature of spin that was not taken into account in \cite{AVS-cor}. Note that in \cite{AVS-cor} only the correlations in close pairs are taken into account. This approximation is insufficient for the description of the OMAR shape $H^2/(|H|+H_0)^2$ that appears in the sample on Fig. \ref{fig:specsample} only when long-range pair correlations are included into the theory.

The author does not know about any way to make Monte-Carlo simulations where electron spin is allowed to have arbitrary direction (instead of only ``up'' and ``down'') and take into account quantum spin correlations. Any wavefunction of a single spin $1/2$ is the eigenfunction of some operator of spin projection $\hat{s}_{\alpha} = c_x \hat{s}_x + c_y \hat{s}_y + c_z \hat{s}_z$, where  $c_x^2 + c_y^2 + c_z^2 = 1$. Therefore it is tempting to describe the electron spins $1/2$ as classical units vectors. However, this model cannot describe the real quantum statistics of spins. Consider for example the scalar product of two spins averaged over some ensemble $\langle {\bf s}_i {\bf s}_j \rangle = \overline{s_{i}^x s_j^x} + \overline{s_{i}^y s_j^y} + \overline{s_{i}^z s_j^z}$. In classical statistics of unit vectors $-1 \le \langle {\bf s}_i {\bf s}_j \rangle \le 1$. The value $1$ describes the vectors that always have the same direction. The value $-1$ corresponds to the vectors that have opposite directions. In quantum mechanics $-3 \le \langle {\bf s}_i {\bf s}_j \rangle \le 1$. The value $-3$ corresponds to the singlet state of the spins. In the present study operator $\hat{s}_i^z$ was selected to have eigenvalues $1$ and $-1$, however, it corresponds to the actual spin $1/2$. In quantum mechanics the eigenvalues of the operator $\hat{l}^2$  of squared angular momentum  are equal to $l(l+1)$. It the case of the angular momentum $1/2$ the only existing eigenvalue is $3/4$. Therefore the operator $(\hat{\bf s}_i)^2 = \hat{s}_i^z\hat{s}_i^z + \hat{s}_i^x\hat{s}_i^x + \hat{s}_i^y\hat{s}_i^y$ is always equal to 3. When two spins are in the singlet state their total spin is equal to zero $(\hat{\bf s}_{i} + \hat{\bf s}_j)^2 = 0$. The product $\hat{\bf s}_i \hat{\bf s}_j$ in this case is well defined and is equal to ${\bf s}_i {\bf s}_j = [(\hat{\bf s}_{i} + \hat{\bf s}_j)^2 - (\hat{\bf s}_i)^2 - (\hat{\bf s}_i)^2]/2 = -3$. When the spins are in a triplet state, the total momentum of the system is equal to $l=1$ and $(\hat{\bf s}_{i} + \hat{\bf s}_j)^2 = 4l(l+1) = 8$. It leads to ${\bf s}_i {\bf s}_j = 1$. Therefore the quantum spin statistics is different from the statistics of classical vectors, although it can be described with the classical values: averaged spin polarizations and their correlations.
The Monte-Carlo calculations at least with a naive description of electron spins cannot be used to quantitatively calculate the magnetoresistance related to the spin correlations and act as ``an arbiter'' for CKE approach.


Although the properties of transport in organic semiconductors are studied for some time they are not completely understood. The low-field mobility in organic semiconductors is often extremely small $\sim 10^{-8} - 10^{-6}cm^2/Vs$ \cite{LM-1,LM-rev,LM-2}. However, these small values can be related to the long-range correlations of electrostatic potential produced by the unscreened molecular dipoles \cite{novikov1995,dunlap,novikov2012}. There is an evidence that at small length-scales the electron mobility can be much higher \cite{Drew}. The provided results show that some information about short-range mobility can be obtained from the measurements of OMAR. The lineshape of organic magnetoresistance depends on hopping rates. However, it is not related to the time for an electron to cross the macroscopic sample. What is important is the time $\tau_{sep}$ that is required for two spins to become separated with sufficient distance that prevents their meeting before their spin correlation relaxes.

A theoretical estimate of $\tau_{sep}$ can be possible in the framework of complex numerical modeling of organic semiconductors similar to the one made in \cite{Masse}. However, even the analysis of the provided simplified models can give some hints on $\tau_{sep}$ and nature of the short-range transport. When $\tau_{sep}$ is small compared to the period of precession in hyperfine field $\tau_{sep} h_{hf} \ll 1$ OMAR is suppressed. When $\tau_{sep} h_{hf} \gtrsim 1$ due to overall slow rate of hopping OMAR can be relatively strong $\sim 10\%$ as shown on Fig. \ref{fig:difw0} and its lineshape is described by Lorentzian or by Eq.~(\ref{R4}). When some of the hops are fast but $\tau_{sep} h_{hf} \gtrsim 1$ due to bottlenecks with slow hops as in the sample shown in Fig.~\ref{fig:specsample} the size of OMAR is still $\sim 10\%$  but is shape is close to $H^2/(|H| + H_0)^2$.

The present study deals only with the most simple model with large Hubbard energy and small applied electric field. In principle, it is possible to generalize CKE theory to include other cases. In \cite{CKE0} the far from equilibrium CKE that can be applied for high electric fields are derived for charge correlations only. In \cite{AVS-cor} the situation with arbitrary Hubbard energy is considered for close-range pair correlations with the semi-classical spin model. It is shown in \cite{AVS-cor,CKE0} that the discussed generalizations make the theory much more complex. The present work shows that different lineshapes of OMAR appear even in the simplest model due to the different properties of short-range transport.

\section{Conclusions}
\label{s:conc}

The system of correlation kinetic equations (CKE) is derived for the spin correlations in materials with hopping transport with large Hubbard energy for a small applied electric field. The spins are assumed to be conserved in the hopping process and can rotate around on-site hyperfine fields. The spin degrees of freedom are described with quantum mechanics as averaged products of spin operators. The derived CKE approach allows to describe the bipolaron mechanism of OMAR. It is shown that the shape of the magnetic field dependence of resistivity contains information about short-range electron transport. Different statistics of hopping rates lead to different OMAR lineshapes including the empirical laws $H^2/(H^2 + H_0^2)$ and $H^2/(|H| + H_0 )^2$ that are often used to describe experimental data.

The author is grateful for many fruitful discussions to Y.M. Beltukov, A.V. Nenashev, D.S. Smirnov, V.I. Kozub and V.V. Kabanov.
The work was supported by the Foundation for the Advancement of Theoretical Physics and Mathematics ``Basis''.
A.V.S. acknowledges support from Russian Foundation for Basic Research (Grant No. 19-02-00184).

\appendix
\section{Derivation of $(d/dt)_{ij}$ term in kinetic equations}
\label{Ap1}

In this appendix I derive the term $(d/dt)_{ij} \overline{s}_I^P$ in the kinetic equation. It is supposed that sites $i$ and $j$ are included into set $I$. For definiteness the site $i$ is considered to be an $A$-type site and $j$ to have type $B$. $\overline{s}_I^P$ can be expressed as quantum mechanical average of the operator
\begin{equation}
 \overline{s}_I^P = \left\langle \hat{s}_I^P \right\rangle =  \left\langle \hat{s}_i^p \hat{s}_j^q \hat{s}_{I'}^{P'} \right\rangle.
\end{equation}
Here when the index $p$  corresponds to $x$,$y$ and $z$, $\hat{s}_i^p$ is the spin polarization operator $\hat{s}_i^{\alpha} = a_{i,n}^+ (\sigma_\alpha)_{nm} a_{i,m}$. $\sigma_\alpha$  is a Pauli matrix. $a_{i,n}^+$ and $a_{i,m}$ are the creation and destruction operators for electron on site $i$. Indexes $n$ and $m$ correspond to spin states $\uparrow$ or $\downarrow$. The operator $\hat{s}_i^0$ is the operator of single occupation and can be expressed as follows
\begin{equation}\label{A-s0}
\hat{s}_i^0 = a_{i\uparrow}^+ a_{i\uparrow} + a_{i\downarrow}^+ a_{i\downarrow} - 2 a_{i\uparrow}^+ a_{i\uparrow} a_{i\downarrow}^+ a_{i\downarrow}.
\end{equation}
The expression (\ref{A-s0}) is valid for any type of site, therefore $\hat{s}_{j}^0$ can be expressed in a similar way.

The transitions between sites $i$ and $j$ can be described by the operator $\widehat{H}_{ij}$ (it is the term in Hamiltoinan related to these transitions).
\begin{equation}
\widehat{H}_{ij} = t_{ij} \widehat{\tau}_{ij} \Phi_{ij} + t_{ji} \widehat{\tau}_{ji}\Phi_{ji}, \quad \tau_{ij} = a_{i\uparrow}^+a_{j\uparrow} + a_{i\downarrow}^+a_{j\downarrow}.
\end{equation}
The operator $\Phi_{ij}$ describes the interaction with phonons that appears in the transition term of Hamiltonian $\widehat{H}_{ij}$ after the polaron transformation \cite{BryksinBook}.

\begin{widetext}

With the approximations corresponding to hopping transport the time derivative of operator $\hat{s}_I^P$ can be expressed as follows
\begin{equation} \label{A-com1}
\left(\frac{d}{dt}\right)_{ij} \hat{s}_I^P = -\frac{1}{\hbar^2} \int_{-\infty}^{t} \left\langle \left[ \widehat{H}_{ij} (t), \left[\widehat{H}_{ij} (t') \hat{s}_I^P(t)  \right] \right] \right\rangle_{ph} dt'.
\end{equation}
Here square brackets denote commutator $[\widehat{A},\widehat{B}] = \widehat{A}\widehat{B} - \widehat{B}\widehat{A}$. $\left\langle ...\right\rangle_{ph}$ means the averaging over phonon variables with equilibrium distribution of phonons. Note that even with the simplifications made eq.~(\ref{A-com1}) contains not only hopping terms but also terms corresponding to quantum mechanical perturbation of electron wavefunctions on sites $i$ and $j$ and to exchange interaction between electrons that is neglected in our study. In further calculations I keep only the terms related to hopping process that are proportional to the hopping rates
\begin{equation}
W_{ij} = \frac{1}{\hbar^2} |t_{ij}|^2 \left\langle \int_{-\infty}^0 \Phi_{ji}(t) \Phi_{ij}(0) e^{(i/\hbar)(\varepsilon_j - \varepsilon_i)t} +
\Phi_{ji}(0)\Phi_{ij}(t) e^{(i/\hbar)(\varepsilon_i-\varepsilon_j)t} dt
 \right\rangle_{ph},
\end{equation}
\begin{equation}
W_{ji} = \frac{1}{\hbar^2} |t_{ij}|^2 \left\langle \int_{-\infty}^0 \Phi_{ij}(t) \Phi_{ji}(0) e^{(i/\hbar)(\varepsilon_i - \varepsilon_j)t} +
\Phi_{ij}(0)\Phi_{ji}(t) e^{(i/\hbar)(\varepsilon_j-\varepsilon_i)t} dt
 \right\rangle_{ph}.
\end{equation}
The term in $(d/dt)_{ij} \widehat{s}_I^P$ proportional to $W_{ji}$ and related to hops $i\rightarrow j$ is equal to $- W_{ji}(
\widehat{\tau}_{ij}\widehat{\tau}_{ji}\widehat{s}_I^P + \overline{s}_I^P\widehat{\tau}_{ij}\widehat{\tau}_{ji}
)/2 $. The term related to $j \rightarrow i$ hops is $W_{ij} \widehat{\tau}_{ji} \widehat{s}_I^P \widehat{\tau}_{ij}$. Here I took into account that site $i$ cannot be double-occupied and site $j$ cannot have zero electrons.

The following computation is quite cumbersome but straightforward operator algebra. There are two useful relations that make it simpler
\begin{equation}
\hat{\tau}_{ij}\hat{\tau}_{ji} = \frac{\hat{s}_i^{0}\hat{s}_j^{0} - \sum_{\alpha} \hat{s}_i^{\alpha}\hat{s}_j^{\alpha}  }{2},
\quad
\hat{s}_{i}^{\alpha}\hat{s}_i^{\beta} = \delta_{\alpha\beta} \hat{s}_i^{0} + i\epsilon_{\alpha\beta\gamma} \hat{s}_i^{\gamma},
\end{equation}
where $\epsilon_{\alpha\beta\gamma}$ is Levi-Civita symbol.

The operator algebra yields
\begin{equation}\label{A-kin-s-00}
\left( \frac{d}{dt} \right)_{ij} \hat{s}_i^0\hat{s}_j^0 \hat{s}_{I'}^{P'} = 2W_{ij} \hat{n}_{j\uparrow} \hat{n}_{j\downarrow} (1 - \hat{n}_{i\uparrow})(1 - \hat{n}_{i\downarrow}) \hat{s}_{I'}^{P'}  -
 \frac{W_{ji}}{2} \left(\hat{s}_i^0 \hat{s}_j^0 - \sum_{\alpha} \hat{s}_i^{\alpha}\hat{s}_j^{\alpha} \right)\hat{s}_{I'}^{P'},
\end{equation}
\begin{equation} \label{A-kin-s-0a}
\left( \frac{d}{dt} \right)_{ij} \hat{s}_i^0\hat{s}_j^\alpha \hat{s}_{I'}^{P'} = \frac{W_{ji}}{2}
\left(\hat{s}_i^{\alpha}\hat{s}_j^0 - \hat{s}_i^{0}\hat{s}_j^\alpha\right) \hat{s}_{I'}^{P'},
\end{equation}
\begin{equation}\label{A-kin-s-a0}
\left( \frac{d}{dt} \right)_{ij} \hat{s}_i^\alpha \hat{s}_j^0 \hat{s}_{I'}^{P'} = \frac{W_{ji}}{2}
\left(\hat{s}_i^{0}\hat{s}_j^\alpha - \hat{s}_i^{\alpha}s_j^0\right) \hat{s}_{I'}^{P'},
\end{equation}
\begin{multline}\label{A-kin-s-ab}
\left( \frac{d}{dt} \right)_{ij} \hat{s}_i^\alpha \hat{s}_j^\beta \hat{s}_{I'}^{P'} =
 -2W_{ij} \delta_{\alpha\beta} \hat{n}_{j\uparrow} \hat{n}_{j\downarrow} (1 - \hat{n}_{i\uparrow})(1 - \hat{n}_{i\downarrow}) \hat{s}_{I'}^{P'}  +
 \frac{W_{ji}}{2} \delta_{\alpha\beta} \left(\hat{s}_i^0 \hat{s}_j^0 - \sum_{\alpha} \hat{s}_i^{\gamma}\hat{s}_j^{\gamma} \right)\hat{s}_{I'}^{P'} - \\
 \frac{W_{ji}}{2} \left(\hat{s}_i^{\alpha}\hat{s}_j^{\beta} - \hat{s}_i^{\beta} \hat{s}_j^{\alpha} \right)\hat{s}_{I'}^{P'}
\end{multline}
The quantum mechanical averaging of these equations leads to eqs.~(\ref{kin-s-00}-\ref{kin-s-ab}) from the main text.

\end{widetext}

\end{document}